\documentclass[a4paper, 12pt]{article}
\usepackage{amsmath,amssymb,amsfonts,amsthm, mathtools}
\usepackage{fullpage,url, hyperref}
\usepackage{longtable, threeparttable, booktabs, graphicx, float}
\usepackage{times, txfonts, color}
\usepackage{titlesec}
\usepackage{algorithmic}
\usepackage[linesnumbered,ruled]{algorithm2e}
\usepackage{subfigure}
\usepackage{mathtools}
\usepackage{mathrsfs}
\usepackage{natbib}
\usepackage{caption}
\usepackage{setspace}

\theoremstyle{definition}
\newtheorem{definition}{Definition}
\newtheorem{remark}{Remark}

\SetCommentSty{mycommfont}

\date{}

\begin{document}

 \title{\bf Granger Causality for Mixed Time Series Generalized Linear Models: A Case Study on Multimodal Brain Connectivity}
 \author{Luiza S.C. Piancastelli$^1$\footnote{\texttt{email}: \href{mailto:luiza.piancastelli@ucd.ie}{luiza.piancastelli@ucd.ie}},\,\, Wagner Barreto-Souza$^1$\footnote{\texttt{email}: \href{mailto:wagner.barreto-souza@ucd.ie}{wagner.barreto-souza@ucd.ie}},\,\,\, Norbert J. Fortin$^{2,3}$\footnote{\texttt{email}: \href{mailto:norbert.fortin@uci.edu}{norbert.fortin@uci.edu}}, \\ \, Keiland W. Cooper$^{2,3}$\footnote{\texttt{email}: \href{mailto:kwcooper@uci.edu}{kwcooper@uci.edu}} \,\, and\, Hernando Ombao$^{4,5}$\footnote{\texttt{email}: \href{mailto:hernando.ombao@kaust.edu.sa}{hernando.ombao@kaust.edu.sa}}\\
\small \it $^1$School of Mathematics and Statistics, University College Dublin, Dublin 4, Republic of Ireland\\
\small \it $^2$Center for the Neurobiology of Learning and Memory, University of California, Irvine, CA, USA \\
\small \it $^3$Department of Neurobiology and Behavior, University of California, Irvine, CA, USA \\
\small \it $^4$Statistics Program, King Abdullah University of Science and Technology, Thuwal, Saudi Arabia \\
\small \it $^5$Neuroscience-AI Laboratory, King Abdullah University of Science and Technology, Thuwal, Saudi Arabia}

 \maketitle

\begin{abstract}
This paper is motivated by studies in neuroscience experiments to understand interactions between nodes in a brain network using different types of data modalities that capture different distinct facets of brain activity. 
To assess Granger-causality, we introduce a flexible framework through a general class of models that accommodates mixed types of data (binary, count, continuous, and positive components) formulated in a generalized linear model (GLM) fashion. Statistical inference for causality is performed based on both frequentist and Bayesian approaches, with a focus on the latter. Here, we develop a procedure for conducting inference through the proposed Bayesian mixed time series model. By introducing spike and slab priors for some parameters in the model, our inferential approach guides causality order selection and provides proper uncertainty quantification. The proposed methods are then utilized to study the rat spike train and local field potentials (LFP) data recorded during the olfaction working memory task. 
The proposed methodology provides critical insights into the causal relationship between the rat spiking activity and LFP spectral power.  Specifically, power in the LFP beta band is predictive of spiking activity 300 milliseconds later, providing a novel analytical tool for this area of emerging interest in neuroscience and demonstrating its usefulness and flexibility in the study of causality in general. 
\end{abstract}


\section{Introduction}

\subsection{Scientific Aims}\label{intro:data}

\begin{figure}[h!]
	\centering
	\includegraphics[angle=-90,width=.9\textwidth]{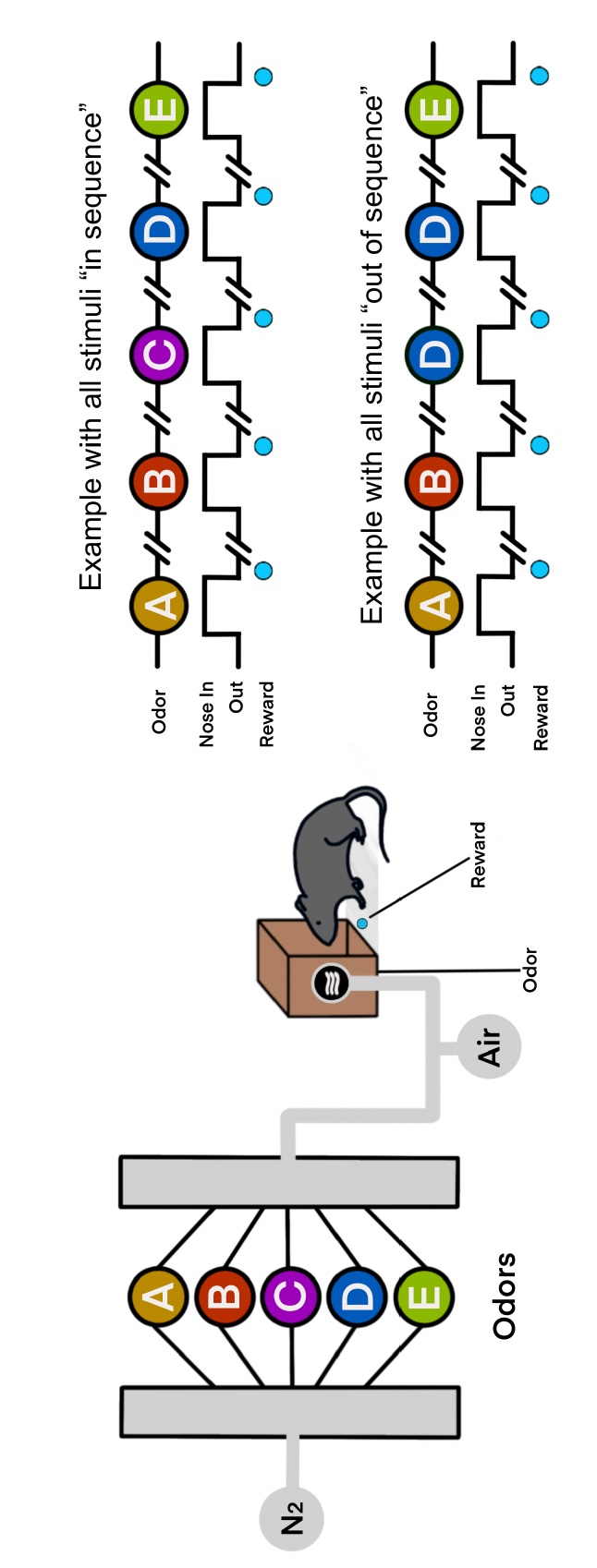}	
	\caption{Rats received multiple sequences of five odors (left; odors A, B, C, D, E). The animals were required to correctly identify whether the odor was presented ``in sequence" (top right; by holding its nose in the port for $\sim$1.2 s, when an auditory signal is delivered) or ``out of sequence" (bottom right; by withdrawing its nose before the signal) to receive a water reward.}\label{fig:ratstudy}
\end{figure}

\begin{figure}[h!]
    \centering  
    \includegraphics[width = 0.8\linewidth]{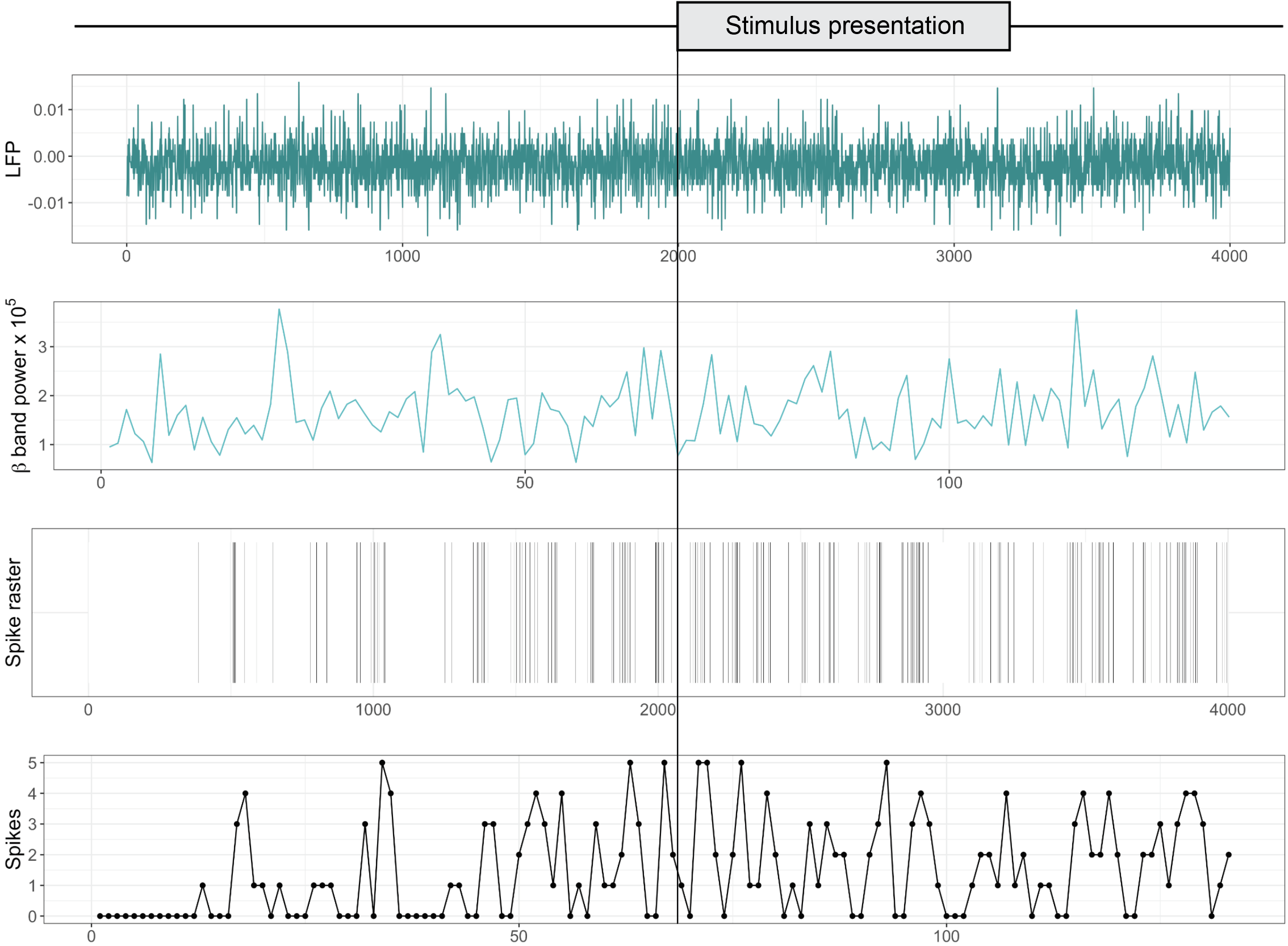}
    \caption{
    Brain activity data from the rat hippocampus during the performance of a complex sequence memory task. The activity from one electrode is shown during a representative trial (one stimulus presentation). Row 1 shows the raw (unfiltered) local field potential (LFP) activity. Row 2 displays the corresponding $\beta$ band power from the LFP in Row 1. Row 3 shows the spiking activity. Row 4 shows the corresponding spike counts, derived from Row 3, within non-overlapping windows of 30 milliseconds.}
    \label{fig:lfp_spike_fig}
\end{figure}

In this paper, we develop a novel statistical model for identifying cross-dependency in a brain network through multiple types of recorded data (multi-modal) that capture different facets of brain activity in rats. This is motivated by the goal of the co-author’s Fortin research (Neurobiology Laboratory, UC Irvine) which is to investigate the ability of animals to remember the specific order (or sequence) of occurrence of events and also to understand the role of the hippocampus region in this capacity. The Fortin lab conducted an experiment in which neural activity - both neuronal spike train and local field potentials (LFPs) -  were recorded in the hippocampus of rats as they performed a complex non-spatial sequence memory task which is similar to paradigms used in humans \citep{TAllen2014}. 
Intracranial electrodes were implanted in the hippocampus and both spiking activity and local field potential (LFP) activity were recorded while performing the task. The rats were trained to recognize a sequence of five different odors (A = Lemon, B = Rum, C = Anise, D = Vanilla, E = Banana). A trial (i.e., a single odor presentation within the sequence) is labeled as ``in sequence" (InSeq) if the odor is presented in the correct sequence position (e.g., AB\underline{C}\ldots); otherwise, the trial is labeled as ``out of sequence" (OutSeq; e.g., AB\underline{D}\ldots), as illustrated in Figure \ref{fig:ratstudy}. 
Hippocampal spiking activity time series and LFP recordings are 
displayed in Figure~\ref{fig:lfp_spike_fig}.
The particular aim that we address in this paper is to characterize the spiking activity of neurons and how these may be activated or 
inhibited by band-specific spectral activity as measured by the LFPs. 

While spiking activity represents discrete events (the specific milliseconds at which individual neurons fired action potentials), LFPs represent continuous and real-valued signals (fluctuating around 0) reflecting the summation of voltage signals near the electrode tips. Given the often oscillatory nature of LFP signals, they are typically analyzed in the frequency or spectral domain \citep{fieomb2016,gaoetal2020,graetal2022}. While the current standard in the field is to focus on analyzing (or decoding) the information contained in the spiking activity of neuronal ensembles, doing so is technically challenging, may result in unpredictable yields, and also ignores potentially meaningful information contained in the LFP \citep{panetal2009,zhouetal2016}. In contrast, while recording LFP activity is methodologically less complex than spiking information, developing analytical tools to extract meaningful information from LFP has proved challenging. Moreover, there is an increasing interest in the joint analysis of these heterogeneous (mixed) data types. 

To help bridge this gap, this paper aims to quantify the influence of spiking activity on LFP and vice-versa.  We analyze LFPs via periodograms computed on established frequency bands \citep{ombpin2022}.
A periodogram expresses the LFP time series as the linear combination of Fourier complex exponentials (or cosine and sine waves): $$Z_t = \alpha_0 + \sum_{j=1}^{(n-1)/2}  \left(  a_j \cos(2\pi tj/n) + b_j \sin(2\pi tj/n) \right),$$
where $n$ denotes the time series sample size; for instance, see \cite{shusto2017}. The coefficients $a_j, b_j$ are easily obtained via the discrete Fourier transform (DFT) and are used to identify the dominant frequencies of the LFP (i.e., the oscillations with the largest amplitudes account for most variation in the signal). The traditional analysis of electrophysiological signals such as LFPs computes the spectral estimates at the following frequency bands $\delta: [1-4] Hz$, $\theta: (4-8] Hz$, $\alpha: (8-13] Hz$, $\beta: (13-30] Hz$ and $\gamma: (30-100] Hz$. These spectral estimates are derived by aggregating periodograms across each frequency band. Note that these bands have been associated with different cognitive and information processes.

We then perform a comparison of the spectral band powers by computing its value at a fixed size window for all possible such intervals in a trial. These values are then grouped according to the following rule: \textit{is there a spike in the time point that succeeds that window?} An exploratory analysis using two-sample tests suggests that the $\beta$ power (20 - 40 Hz) is a potential predictor of spiking activity, which will be formally tested with the model that will be developed in this paper. 
To this end, the original continuous-valued LFP trajectory $Z_t$ is transformed into a positive-valued series by using the $\beta$ power at non-overlapping windows of 30 milliseconds (which contains 30 discrete time points). The window size is chosen in such a way that it is sufficiently large to capture the signal strength while still providing adequate temporal resolution. Note that one can use smaller windows to produce better time localization (which is ideal for more time-reactive measurements), but this comes at the cost of having fewer observations and hence reducing discrimination of the signal strength at the band. 

One of our aims is to study the Granger-causality between spectral activity of LFPs (strictly positive-valued time series) and spiking activity as characterized by the number of spikes (count time series). While there is a plethora of models for binary/count time series and also many established methods for continuous-valued time series, there is a lack of models and methods for analyzing mixed types of data. In our specific analysis, the spike rate trajectory is constructed by aggregation of the originally binary spike/no-spike indicators in the same windows of the LFP spectral power. This results in continuous positive and integer-valued trajectories of length $n=133$ which are obtained as transformations of the original data that has 4000 time points. 

The study of causality between spikes and LFPs is of emerging interest and has been considered for instance by \cite{hu2016} and \cite{gong2019spike}. In \cite{hu2016}, a Gaussian copula-based approach indicates a strong Granger causality between LFPs and spikes in the LFP $\rightarrow$ spike direction. We also aim at the study of pairwise dependencies between spike trajectories from five different electrodes. While contemporaneous causality between pairs is expected, our goal is to evaluate if this holds between all possible ${5 \choose 2}$ combinations. To address this, we will develop a model that aims to address neuroscientific questions on whether or not lagged effects can meaningfully explain the Granger causality between any bivariate trajectories.

\subsection{Literature Review}
The study on the interactions among time series has been of interest in areas such as finance, economics, and neuroscience since the seminal paper by \cite{gra1969}, where the concept of Granger causality was introduced. For some applications in these areas, for instance, see \cite{sim1980}, \cite{lee1992}, \cite{hongetal2009}, and \cite{setetal2015}. Testing for causality in the presence of nonstationary time series has been proposed by \cite{lietal2014} for bivariate autoregressive processes, while \cite{guoetal2014} proposed an approach for investigating causality based on univariate factor double autoregressive models.

The majority of the approaches for testing Granger causality rely on the linear vector autoregressive (VAR) model, which involves the multivariate normal distribution or some multivariate continuous distribution. However, a major limitation of VAR models is that the different components of time series data may possess different attributes (e.g., binary, counts, and positive continuous). Granger causality problem involving count time series has been recently addressed in the literature. \cite{cheandlee2017} investigated the causal relationship between climate and criminal behavior based on univariate integer-valued generalized autoregressive conditional heteroscedasticity (INGARCH) models, which is a proper approach to handle count time series. Bivariate INGARCH models have also been considered to study Granger causality involving counts; for instance, see \cite{leeandlee2019} and \cite{piaetal2023}. \cite{tanetal2021} proposed a test for dealing with multivariate categorical time series based on the mixture transition distribution model. For a review accounting for the most important developments and recent advances on Granger causality, we recommend the paper by \cite{shofox2022}.

The aforementioned models are not able to address the specific questions on brain causality inferred from observed signals of different modality types, for instance, the number of spikes in a local time window (count) and the spectral power at various frequency bands in the same local time window (positive-valued). The blind use of models that ignore the correct nature of data can produce misleading results. This is our motivation to introduce a proper class of bivariate time series allowing for mixed components, as discussed in the next subsection.

\subsection{Contributions}
The main contribution of this paper is the framework for proper Granger-causality tests for multi-modal brain activity data to address the scientific questions discussed in Subsection \ref{intro:data}. We propose a novel family of bivariate time series generalized linear models for analyzing different types of data (multimodal) such as binary, counts, continuous, and positive-valued time series. To the best of our knowledge, this is the first formal exploration of Granger-causality under a model with mixed data. In addition, our proposed model is able to handle other challenges, namely, 
(i) modeling non-linear interactions between various data types; (ii) simultaneous mean and variance Granger-causality \citep{guoetal2014}; (iii) allowance for contemporaneous causality. 

Statistical inference for our time series families and Granger causality test are performed based on both frequentist and Bayesian approaches, with a focus on the latter. In addition to readily an appealing interpretation to practitioners, model selection is further developed under the Bayesian paradigm. More specifically, an inferential approach with spike and slab priors is carefully designed to address the causality order selection. Causality order selection here is the study of synchrony between the LFP power and the spiking activity, which allows us to answer the following \textit{How long does it take for an observed increase in activity at the LFP band to cause a spike?} This main point of interest to neuroscientists is answered elegantly under the spike and slab approach that provides proper uncertainty quantification. It is worth mentioning that many challenges stated in the recent survey by \cite{shofox2022} are addressed by our methodology.

\subsection{Organization of the Paper}
The paper is organized as follows. In Section \ref{sec:model}, we introduce some notation and define the class of mixed time series generalized linear models. Estimation of parameters is addressed in Section \ref{sec:inference} under both frequentist and Bayesian perspectives, with emphasis on the latter. Granger causality testing is covered in Section \ref{sec:GC} based on the time series GLM. In particular, we develop a method to select/estimate the causality order using spike and slab priors, including uncertainty quantification in Subsection \ref{sec:spike_slab}. Two studies on Granger causality for brain activity data are presented in Section \ref{sec:braindataapp}. Concluding remarks and future research are discussed in Section \ref{sec:discussion}.

\section{Model definition}\label{sec:model}

We begin by introducing some notation. A random variable $Y$ is said to be a member of the exponential family (EF) if its density/probability function assumes the form
\begin{eqnarray*}
f(y)=\exp\left\{\dfrac{y\zeta-b(\zeta)}{\phi}+c(y;\phi)\right\}, \quad y\in\mathcal S,	
\end{eqnarray*}	 
where $\mathcal S\subset\mathbb R$ is the support of the distribution, $b(\cdot)$ is assumed to be a continuous three-times differentiable function, $c(\cdot,\cdot)$ is some function mapping $\mathcal S\times\mathbb R^+$ into $\mathbb R$, and $\phi>0$. The moment generating function of $Y$, in this case, is given by 
\begin{eqnarray}\label{mgf}
\Psi(t)\equiv E(e^{tY})=\exp\left\{\dfrac{b(\zeta+\phi t)-b(\zeta)}{\phi}\right\},
\end{eqnarray}
for $t$ belonging to some interval containing the value zero. The mean and variance of $Y$ are respectively given by $\mu\equiv E(Y)=d b(\zeta)/d\zeta\equiv \dot b(\zeta)$ and $\mbox{Var}(Y)=\phi d^2b(\zeta)/d\zeta^2\equiv \phi \ddot b(\zeta)\equiv\phi V(\mu)$, where $V(\cdot)$ is called by variance function. The inverse of the function $\dot b(\zeta)$ (first derivative of $b(\cdot)$) is denoted by $\eta(\cdot)$. Further, we denote $Y\sim\mbox{EF}(\mu,\phi)$, with $\mu$ and $\phi$ being mean and dispersion parameters, respectively.

We now introduce the bivariate causal EF aiming at the modelling of instantaneous causality for our final model construction. 

\begin{definition} (Bivariate causal EF)
A random vector $(Y_1,Y_2)$ follows a bivariate causal exponential family if $Y_1\sim\mbox{EF}^{(1)}(\mu_1,\phi_1)$ and $Y_2|Y_1=y_1\sim\mbox{EF}^{(2)}(\mu_2h_\rho(y_1),\phi_2)$, where $h_\rho(\cdot)$ is some real function mapping into $\mathcal S$ satisfying $E\left(|h_\rho(Y_1)|\right)<\infty$ and $h_{0}(y)=1$, with $\rho\in\mathbb R$ being a parameter controlling the dependence between $Y_2$ and $Y_1$. The superscripts ``(1)" and ``(2)" used in the exponential families aim to make clear that they are not necessarily the same distribution. The functions $b$'s and $c$'s associated with EFs will be made explicitly depending on these superscripts.
\end{definition}

\begin{remark}
	Under the bivariate causal EF model, independence between $Y_2$ and $Y_1$ is obtained by taking $\rho=0$. Regarding the function $h_\rho(\cdot)$, a practical choice is $h_\rho(y)=\exp(\rho y)$ when the support of $Y_2$ is $\mathbb R$, $\mathbb R^+$ or $\mathbb N_0$, and $h_\rho(y)=2\dfrac{\exp(\rho y)}{1+\exp(\rho y)}\equiv 2\mbox{logit}^{-1}(\rho y)$, when the support is $[0,1]$, $(0,1)$ or $\{0,1,\ldots,m\}$, for $m \in\mathbb N$.
\end{remark}


Let $\{Y_{1\,t}\}_{t\in\mathbb N}$ and $\{Y_{2\,t}\}_{t\in\mathbb N}$ be two time series. Denote the sigma-algebras $\mathcal F^{(1)}_{t}=\sigma(Y_{1\,t},Y_{1\,t-1},\ldots)$, $\mathcal F^{(2)}_{t}=\sigma(Y_{2\,t},Y_{2\,t-1},\ldots)$, and $\mathcal F^{(1,2)}_t=\sigma(\mathcal F^{(1)}_{t},\mathcal F^{(2)}_{t})$. Our aim is to test if $\{Y_{1\,t}\}_{t\in\mathbb N}$ causes to $\{Y_{2\,t}\}_{t\in\mathbb N}$ in mean \citep{gra1969}, that is
\begin{eqnarray*}
\mbox{Pr}\left(E(Y_{2\,t}|\mathcal F^{(2)}_{t-1})\neq E(Y_{2\,t}|\mathcal F^{(1,2)}_{t-1})\right)>0.
\end{eqnarray*}

Depending on the distribution choice for $\mbox{EF}^{(2)}$, the variance may depend on the mean. As a consequence, in these cases $\{Y_{1\,t}\}_{t\in\mathbb N}$ will Granger-cause $\{Y_{2\,t}\}_{t\in\mathbb N}$ in both mean and variance simultaneously; see \cite{guoetal2014} for simultaneous causality testing in factor double autoregressions. According \cite{graetal1986}, $\{Y_{1\,t}\}_{t\in\mathbb N}$ causes to $\{Y_{2\,t}\}_{t\in\mathbb N}$ in variance if
\begin{eqnarray*}
\mbox{Pr}\left(\mbox{Var}(Y_{2\,t}|\mathcal F^{(2)}_{t-1})\neq \mbox{Var}(Y_{2\,t}|\mathcal F^{(1,2)}_{t-1})\right)>0.
\end{eqnarray*}

We now define the class of time series generalized linear models based on the bivariate causal exponential family discussed above. Tests for Granger-causality will be conducted under this model formulation.

\begin{definition}\label{ts_granger_model}
	The bivariate class of causal time series generalized linear models (Granger-GLMs) is defined by the time series vector $\{(Y_{1\,t},Y_{2\,t})\}_{t\in\mathbb N}$ satisfying $Y_{1\,t}|\mathcal F^{(1)}_{t-1}\sim\mbox{EF}^{(1)}(\mu_{1\,t},\phi_1)$ and $Y_{2\,t}|\mathcal F^{(1,2)}_{t-1}\sim\mbox{EF}^{(2)}(\mu_{2\,t}h_\rho(Y_{1\,t}),\phi_2)$, with
	\begin{eqnarray}
		\nu_{2\,t}&\equiv& g_2(\mu_{2\,t})\equiv \beta_0^{(2)}+\sum_{i=1}^p \beta_i^{(2)}\widetilde Y_{2\, t-i}+\sum_{j=1}^q \alpha_j^{(2)}\nu_{2\, t-j}+\sum_{l=1}^k \gamma_l \widetilde Y_{1\, t-l},\label{eq2_mu2}\\
				\nu_{1\,t}&\equiv& g_1(\mu_{1\,t})\equiv \beta_0^{(1)}+\sum_{i=1}^r \beta_i^{(1)}\widetilde Y_{1\, t-i}+\sum_{j=1}^s \alpha_j^{(1)}\nu_{1\, t-j},\label{eq1_mu1}
	\end{eqnarray}	
where $g_1(\cdot)$ and $g_2(\cdot)$ are link functions assumed to be continuous, invertible, and twice differentiable, with $\widetilde Y_{1\, t}=T_1(Y_{1\, t})$ and $\widetilde Y_{2\, t}=T_2(Y_{2\, t})$ being adequate transformations of the original time series. 
\end{definition}

\begin{remark}
	The transformed time series in (\ref{eq2_mu2}) and (\ref{eq1_mu1}) via $T_1(\cdot)$ and $T_2(\cdot)$ are necessary since we are modelling transformed mean-related parameters. In general, we will consider $T_1(y)=g_1(y)$ and $T_2(y)=g_2(y)$ unless some slight modification is necessary (e.g., in the count case). More specifically, if $Y_{1\, t}$ is a time series of counts and $g_1(y)=\log y$, we take $T_1(y)=\log(y+1)$ because the log-function will not be well-defined at $y=0$. This is considered in the log-linear INGARCH models by \cite{foktjo2011}. The choices for the transformed time series being the link functions or slight modifications of them will keep $\nu_{1\,t}$ and $\nu_{2\,t}$ in the same scales of $\widetilde Y_{1\, t}$ and $\widetilde Y_{2\, t}$, respectively.
\end{remark}

\section{Estimation of parameters}\label{sec:inference}

\subsection{Frequentist approach}

Assume that $\{(y_{1\, t},y_{2\, t})\}_{t=1}^n$ is a realization of a bivariate causal Granger-GLM according to Definition \ref{ts_granger_model}. We begin by discussing the parametric time series GLM approach. Define the parameter vector $\boldsymbol\theta=(\boldsymbol\beta^{(1)}, \boldsymbol\beta^{(2)}, \boldsymbol\alpha^{(1)}, \boldsymbol\alpha^{(2)}, \boldsymbol\gamma,\phi_1,\phi_2,\rho)^\top$, and $l=\max(p,q,r,s,k)$. The likelihood function can be expressed by $L(\boldsymbol\theta)=\prod_{t=l+1}^n f(y_{1\, t}|\mathcal F_{t-1}^{(1)})f(y_{2\, t}|\mathcal F_{t-1}^{(1,2)})$. The log-likelihood function assumes the form $\ell(\boldsymbol\theta)\equiv \ell_1(\boldsymbol\beta^{(1)}, \boldsymbol\alpha^{(1)},\phi_1)+\ell_2(\boldsymbol\beta^{(2)}, \boldsymbol\alpha^{(2)},\boldsymbol\gamma,\phi_2,\rho)$, where 
\begin{eqnarray*}
\ell_1(\boldsymbol\beta^{(1)}, \boldsymbol\alpha^{(1)},\phi_1)=\sum_{t=l+1}^n\left\{\dfrac{y_{1\, t}\eta_1(\mu_{1\,t})-b_1(\eta_1(\mu_{1\,t}))}{\phi_1}+c_1(y_{1\, t};\phi_1)\right\}	
\end{eqnarray*}	
and 
\begin{eqnarray*}
	\ell_2(\boldsymbol\beta^{(2)}, \boldsymbol\alpha^{(2)},\boldsymbol\gamma,\phi_2,\rho)=\sum_{t=l+1}^n\left\{\dfrac{y_{2\, t}\eta_2\left(\mu_{2\,t}h_\rho(y_{1\,t})\right)-b_2\left(\eta_2\left(\mu_{2\,t}h_\rho(y_{1\,t})\right)\right)}{\phi_2}+c_2(y_{2\, t};\phi_2)\right\}.	
\end{eqnarray*}		

\begin{figure}[h!]
	\centering
	\includegraphics[width=0.3\textwidth]{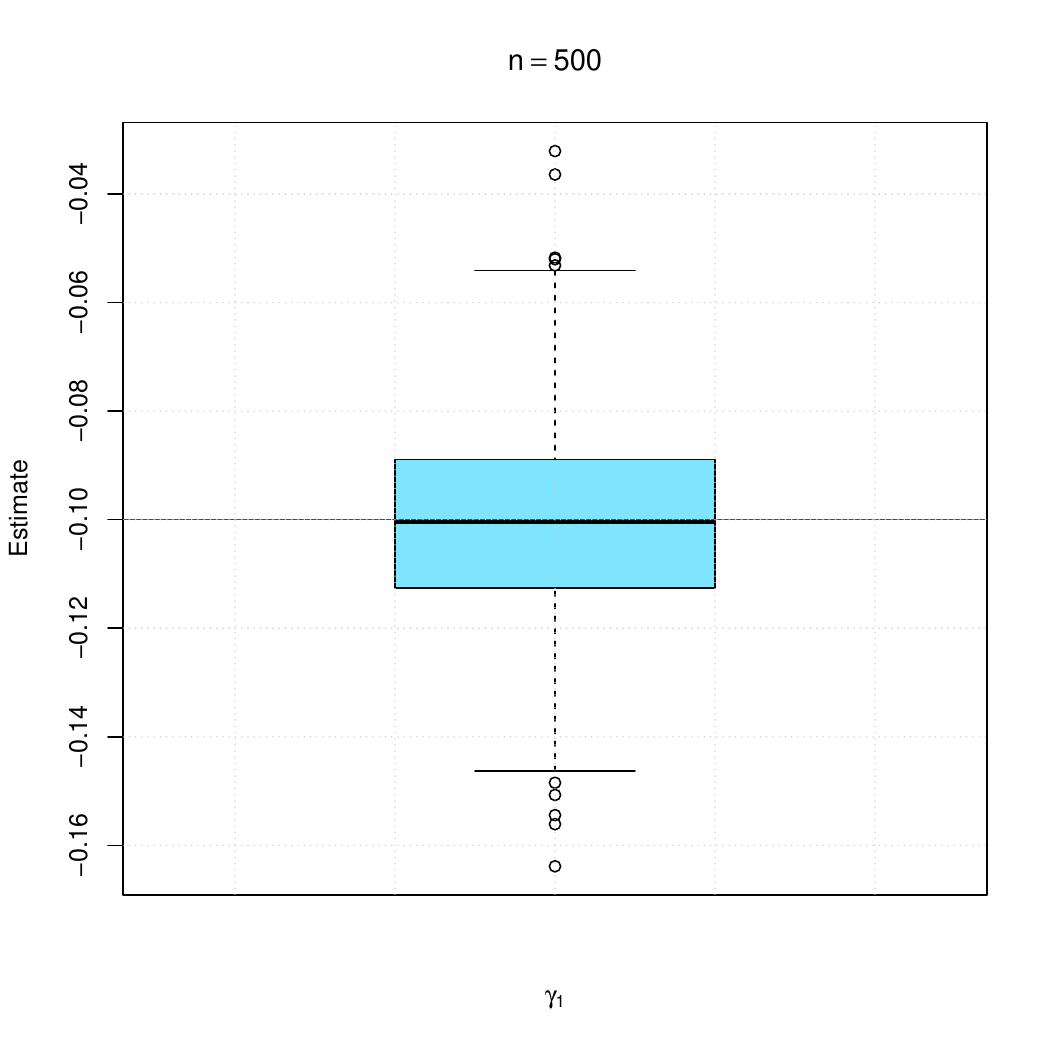}	\includegraphics[width=0.3\textwidth]{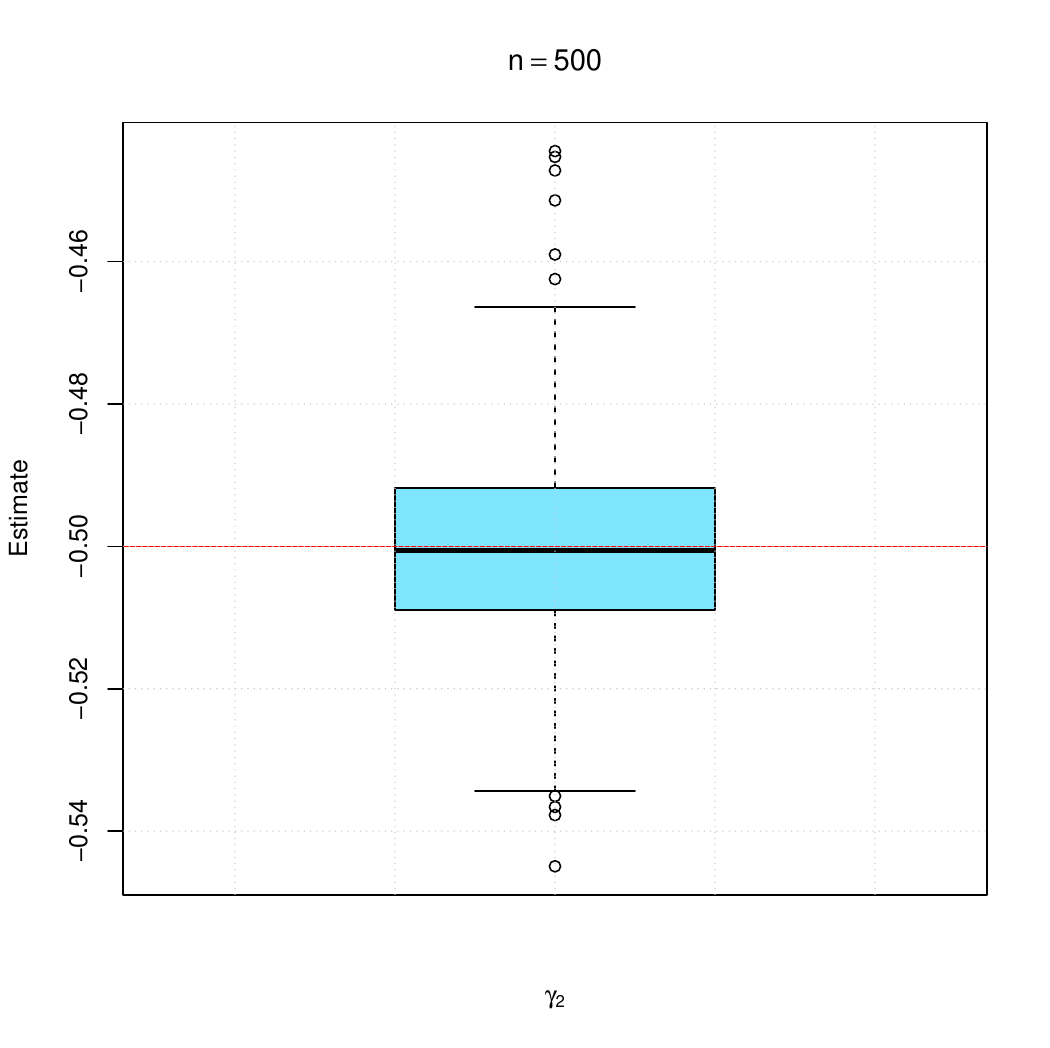}
	\includegraphics[width=0.3\textwidth]{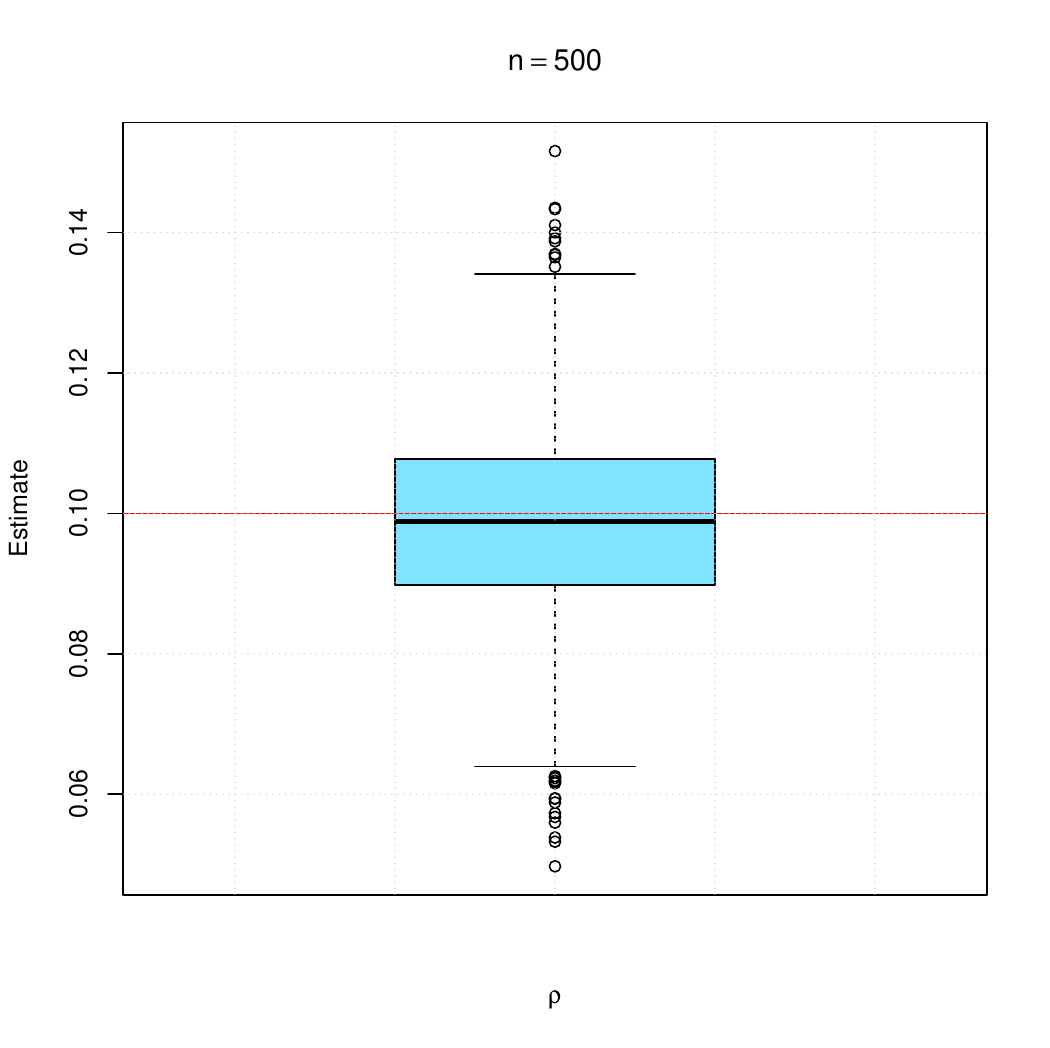}
	\caption{Boxplots of the estimates of the parameters $\gamma_1$, $\gamma_2$, and $\rho$, for $n=500$.}\label{fig:boxplots}
\end{figure}

The maximum likelihood estimator (MLE) of $\boldsymbol\theta$ is given by $\widehat{\boldsymbol\theta}=\mbox{argmax}_{\boldsymbol\theta}\ell(\boldsymbol\theta)$. We now perform a small Monte Carlo simulation to verify the performance of the proposed estimators. Consider the Poisson-Gamma time series model with $T_1(y)=\log y$ and $T_2(y)=\log(y+1)$, $\boldsymbol\beta^{(2)}=(0.2,0.3)^\top$, $\boldsymbol\alpha^{(2)}=(0.2,-0.1)^\top$, $\boldsymbol\gamma=(-0.1,-0.5)^\top$, $\boldsymbol\beta^{(1)}=(0.1,-0.1)^\top$,
$\boldsymbol\alpha^{(1)}=(0.1,0.4)^\top$, $\rho=0.1$, $\phi_1=1$, and sample sizes $n=500,1000,2000$. Table \ref{tab:confI} presents the empirical mean estimates and standard errors of the Poisson-gamma time series model parameters. Figure \ref{fig:boxplots} gives us the boxplots related to the estimates of the parameters $\gamma_1$, $\gamma_2$, and $\rho$, with sample size $n=500$. This small simulation study reveals that the MLEs are performing well as bias and variance decrease as the sample size increases for the model considered, which in some sense is already expected since our model is formulated under GLMs where good property estimators are well-established. Of course, we need to keep in mind that the formulation here is more complex (time-dependent), so a deeper study on the theoretical properties of MLE is needed as discussed in the conclusion section of the paper.

\begin{table}[h!]
	\centering
	\begin{tabular}{c|ccc|ccc}
		\hline
		& \multicolumn{3}{c}{Estimates} & \multicolumn{3}{c}{Stand. errors} \\
		\hline
		$n\rightarrow$ & 500 & 1000 & 2000 & 500 & 1000 & 2000 \\
		\hline
		$\beta^{(2)}_{0}=0.2$ & 0.203 & 0.201 & 0.201 & 0.056 & 0.039 & 0.028 \\
		$\beta^{(2)}_{1}=0.3$ & 0.298 & 0.300 & 0.298 & 0.053 & 0.036 & 0.026 \\
		$\alpha^{(2)}_{1}=0.2$ & 0.201 & 0.200 & 0.201 & 0.056 & 0.038 & 0.027 \\
		$\alpha^{(2)}_{2}=-0.1$ & $-0.101$ & $-0.100$ & $-0.101$ & 0.036 & 0.024 & 0.016 \\
		$\gamma_1=-0.1$ & $-0.100$ & $-0.100$ & $-0.100$ & 0.018 & 0.012 & 0.008 \\
		$\gamma_2=-0.5$ & $-0.500$ & $-0.500$ & $-0.500$ & 0.013 & 0.009 & 0.006 \\
		$\beta^{(1)}_{0}=0.1$ & 0.242 & 0.198 & 0.126 & 0.273 & 0.238 & 0.128 \\
		$\beta^{(1)}_{1}=-0.1$ & $-0.089$ & $-0.093$ & $-0.098$ & 0.045 & 0.035 & 0.022 \\
		$\alpha^{(1)}_{1}=0.1$ & $-0.060$ & $-0.019$ & 0.074 & 0.622 & 0.493 & 0.274 \\
		$\alpha^{(1)}_{2}=0.4$ & 0.017 & $0.160$ & 0.329 & 0.574 & 0.459 & 0.267 \\
		$\phi_1=1$ & 0.991 & 0.998 & 0.999 & 0.056 & 0.039 & 0.028\\
		$\rho=0.1$ & 0.099 & 0.100 & 0.100 & 0.015 & 0.010 & 0.007 \\
		\hline
	\end{tabular}\caption{Empirical mean estimates and standard errors of the parameters under the Poisson-gamma time series model for the sample sizes $n=500,1000,2000$.}\label{tab:confI}
\end{table}

\subsection{Bayesian inference}\label{sec:bayes}

We now explore how to perform inference under a Bayesian perspective, which is one of the main focuses of this paper. A bivariate causal EF model is amenable to standard Bayesian inference once the exponential family distributions $\mbox{EF}^{(1)}$ and $\mbox{EF}^{(2)}$ are specified. In this setting, our interest is to infer the posterior distribution of parameters $\boldsymbol\theta=(\boldsymbol\beta^{(1)}, \boldsymbol\beta^{(2)}, \boldsymbol\alpha^{(1)}, \boldsymbol\alpha^{(2)}, \boldsymbol\gamma,\phi_1,\phi_2,\rho)^\top$ given the observed trajectories $\{(y_{1\, t},y_{2\, t})\}_{t=1}^n$. Let the observed data be denoted by $\boldsymbol{y}_1, \boldsymbol{y}_2$, and $f(\boldsymbol{y}_1, \boldsymbol{y}_2|\boldsymbol{\theta})$ indicate the likelihood model $f(\boldsymbol{y}_1, \boldsymbol{y}_2| \boldsymbol\theta)=\prod_{t=l+1}^n f(y_{1\, t}|\mathcal F_{t-1}^{(1)})f(y_{2\, t}|\mathcal F_{t-1}^{(1,2)})$ that has the mean structure in Definition (\ref{ts_granger_model}). The posterior distribution of $\boldsymbol{\theta}$ with prior $\pi(\boldsymbol\theta)$ is given by

 \begin{equation}\label{eq:posterior}
    p(\boldsymbol{\theta}|\boldsymbol{y}_1, \boldsymbol{y}_2) \propto f(\boldsymbol{y}_1, \boldsymbol{y}_2|\boldsymbol{\theta}) \pi(\boldsymbol{\theta}),
\end{equation}
up to a proportionality constant. In this section, our goal is to stipulate a simple strategy to sample from (\ref{eq:posterior}) using Markov Chain Monte Carlo (MCMC). In MCMC, we collect samples from (\ref{eq:posterior}) by drawing from the full-conditional distributions of the model parameters in such a way that a Markov chain that has a stationary distribution equal to the target (\ref{eq:posterior}) is constructed. Once the chain can be considered to have reached its stationary state, samples are stored and treated as random realizations from the posterior. 

We avail of the hierarchy in our bivariate causal Granger-GLM model to stipulate algorithms that sample $p(\boldsymbol{\theta}_1|\boldsymbol{y}_1)$, then $p(\boldsymbol{\theta}_2|\boldsymbol{y}_1, \boldsymbol{y}_2, \boldsymbol{\theta}_1)$, where we have defined $\boldsymbol{\theta}_1 \equiv (\boldsymbol{\beta}^{(1)}, \boldsymbol{\alpha}^{(1)}, \phi_1)^\top$, which are the parameters pertaining the causal time series, $\boldsymbol{y}_1$ only, and those related to the conditional distribution of $\boldsymbol{y}_2$ given $\boldsymbol{y}_1$ (the caused trajectory) are $\boldsymbol{\theta}_2 \equiv (\boldsymbol{\beta}_2, \boldsymbol{\alpha}_2, \boldsymbol{\gamma}, \rho, \phi_2)^\top$; therefore, $\boldsymbol{\theta} = (\boldsymbol{\theta}_1^\top, \boldsymbol{\theta}_2^\top)^\top$.

Our strategy is to stipulate Metropolis-Hastings (MH) updates so that the algorithms specified in this section generalize to any choices of $\mbox{EF}^{(1)}$ and $\mbox{EF}^{(2)}$. This is explained next in what follows. For instance, sampling the full-conditional distribution of $\boldsymbol{\theta}_1$ parameters means drawing from $p(\theta_1|\boldsymbol{y}_1) \propto f(\boldsymbol{y}_1|\boldsymbol{\theta}_1) \pi(\theta_1)$. If the latter is of known form, Gibbs sampling can be applied and is the most efficient strategy. However, this would require working out explicitly the form of the full-conditional under distinct choices of $\mbox{EF}^{(1)}$ and $\pi(\theta_1)$. In MH, it suffices to be able to write the target up to proportionality, which gives the flexibility of easily changing our distributional assumptions.

A Metropolis-Hastings sampler for $p(\boldsymbol{\theta}_1|\boldsymbol{y}_1)$ builds a Markov Chain that transitions from a current state of $\boldsymbol{\theta}_1$, denoted $\boldsymbol\theta_1^t$, to a candidate $\boldsymbol \theta_1'$ with probability 
\begin{equation}\label{eq:mh_y1}
    p(\boldsymbol\theta_1'|\boldsymbol\theta_1^t) = \min \left\{1, \frac{f(\boldsymbol{y}_1|\boldsymbol{\theta}_1') \pi(\boldsymbol\theta_1') q(\boldsymbol \theta_1^t|\boldsymbol \theta_1') }{f(\boldsymbol{y}_1|\boldsymbol{\theta}_1^t) \pi(\boldsymbol\theta_1^t) q(\boldsymbol \theta_1'|\boldsymbol \theta^t_1)} \right\}.
\end{equation}
The term $q(\boldsymbol \theta_1'|\boldsymbol \theta^t_1)$ denotes the proposal distribution, which is how candidates are generated, i.e. $\boldsymbol \theta_1' \sim q(\cdot|\boldsymbol \theta^t_1)$. Conditioning on $\boldsymbol \theta^t_1$ denotes that the proposed state is generated in some proximity to the current in our framework. Proximity is dictated by a proposal parameter, as it is common in this type of problem.

We adopt single-site moves, which means updating $\boldsymbol \theta^t_1$ one element at a time by drawing from the full-conditional distribution $\pi(\theta_{1,j}|\boldsymbol{\theta}_{1, -j}, \boldsymbol{y}_1)$, where $\boldsymbol{\theta}_{1, -j}$ denotes $\boldsymbol{\theta}_1$ without its $j^{th}$ element. MH transitions can be elaborated in various ways, but it is well known that joint proposals tend to reduce the algorithm's acceptance rate and can also be difficult to calibrate. With univariate proposals, we draw $\theta_{1,j}'$ from $q_j(\cdot| \theta^t_{1,j})$ taking $q_j( \theta_{1,j}'| \theta^t_{1,j})$ to be a Normal with mean $\theta^t_{1,j}$ and variance $\sigma_{1,j}^2$ for real-valued $\theta_{1,j}$. When $j =r + s + 2$, $q_j( \theta_{1,j}'| \theta^t_{1,j}) \equiv \mbox{log-Normal}(\log  \theta^t_{1,j}, \sigma_j)$ so a new state for the dispersion parameter $\phi_1'$ is drawn from a log-normal distribution with median $\theta^t_{1,j}$ and scaling $\sigma_j$. Once $\theta_{1,j}'$ is proposed, the chain moves to the new state with probability (\ref{eq:mh_y1}), in which case $\boldsymbol\theta_1^{(t+1)} = \theta_{1,j}'$. Otherwise $\theta_{1,j}^{(t+1)} = \theta_{1,j}^t$. 

In an equivalent manner, single-site MH moves are used to update the parameters of the caused time series, $\boldsymbol{\theta}_2 \equiv (\boldsymbol{\beta}_2, \boldsymbol{\alpha}_2, \boldsymbol{\gamma}, \rho, \phi_2)$. For $k = 1, \ldots, p+q+k+2$, the full-conditional distribution of $\theta_{2,k}$ given the other elements of $\boldsymbol{\theta}_2$ and the data is 
\begin{equation}\label{eq:mh_y2}
p({\theta}_{2,k}|\boldsymbol{\theta}_{2, -k}, \boldsymbol{\theta}_1, \boldsymbol{y}_1) \propto f(\boldsymbol{y}_1, \boldsymbol{y}_2|\boldsymbol{\theta}_1, \boldsymbol{\theta}_2) \pi(\theta_{2,k}),
\end{equation}
up to proportionality. Given $\boldsymbol{y}_1, \boldsymbol{y}_2$ and the current states of $\boldsymbol{\theta}_1$ and $\boldsymbol{\theta}_2$,  $\theta_{2,k}'$ is drawn from $q_k(\cdot| \theta^t_{2,k})$ and accepted with probability
\begin{equation}\label{eq:mh_y1}
    p(\theta_{2,k}'|\theta_{2,k}^t) = \min \left\{1, \frac{f(\boldsymbol{y}_1, \boldsymbol{y}_2 |\boldsymbol{\theta}_1^t, \boldsymbol{\theta}_2') \pi(\theta_{2,k}') q_k( \theta_{2,k}^t| \theta_{2,k}') }{f(\boldsymbol{y}_1, \boldsymbol{y}_2|\boldsymbol{\theta}_1^t, \boldsymbol{\theta}_2^t) \pi(\theta_{2,k}^t) q_k( \theta_{2,k}'| \theta^t_{2,j})} \right\}.
\end{equation}
For $k \in \{1, \ldots, p +q +k + 1\}$, $q_k( \theta_{2,k}'| \theta^t_{2,j}) \equiv \mathcal{N}(\theta^t_{2,j}, \sigma^2_{2,k})$ are symmetric Gaussian proposals, so $q_k( \theta_{2,k}^t| \theta_{2,k}')/q_k( \theta_{2,k}'| \theta_{2,k}^t) =1$. When $j = p+q+k+2$, the proposal ratio for $\theta_{2,k}\equiv \phi_2$ is $q_k( \phi_2^t| \phi_2')/q_k( \phi_2'| \phi_2^t) = \phi_2'/\phi_2^t$.

Finally, the choice of prior distributions $\pi(\boldsymbol{\theta}_1)$, $\pi(\boldsymbol{\theta}_2)$ completes the proposed methodology to sample (\ref{eq:posterior}). Initially, we take independent uninformative priors, setting real-valued parameters $\boldsymbol{\theta}_{\mathcal{R}} \equiv (\boldsymbol\beta^{(1)}, \boldsymbol\beta^{(2)}, \boldsymbol\alpha^{(1)}, \boldsymbol\alpha^{(2)}, \boldsymbol\gamma,\rho)$ as independent and normally distributed with mean zero and possibly large variances, a priori. In a similar fashion, independent truncated normal distribution on the interval $[0, \infty]$ are set for $\phi_1, \phi_2>0$. The vector of prior variances corresponding to $\boldsymbol{\theta}_1$ and $\boldsymbol{\theta}_2$ is denoted by $\boldsymbol{\tau}^2$. 
A different prior specification will be explored later and is the focus of Subsection \ref{sec:spike_slab}.  Therein, spike and slab forms of $\pi(\boldsymbol{\theta})$ are explored as a tool of causality order selection.

\begin{figure}[h!]
    \centering
    \includegraphics[width = 0.8\linewidth]{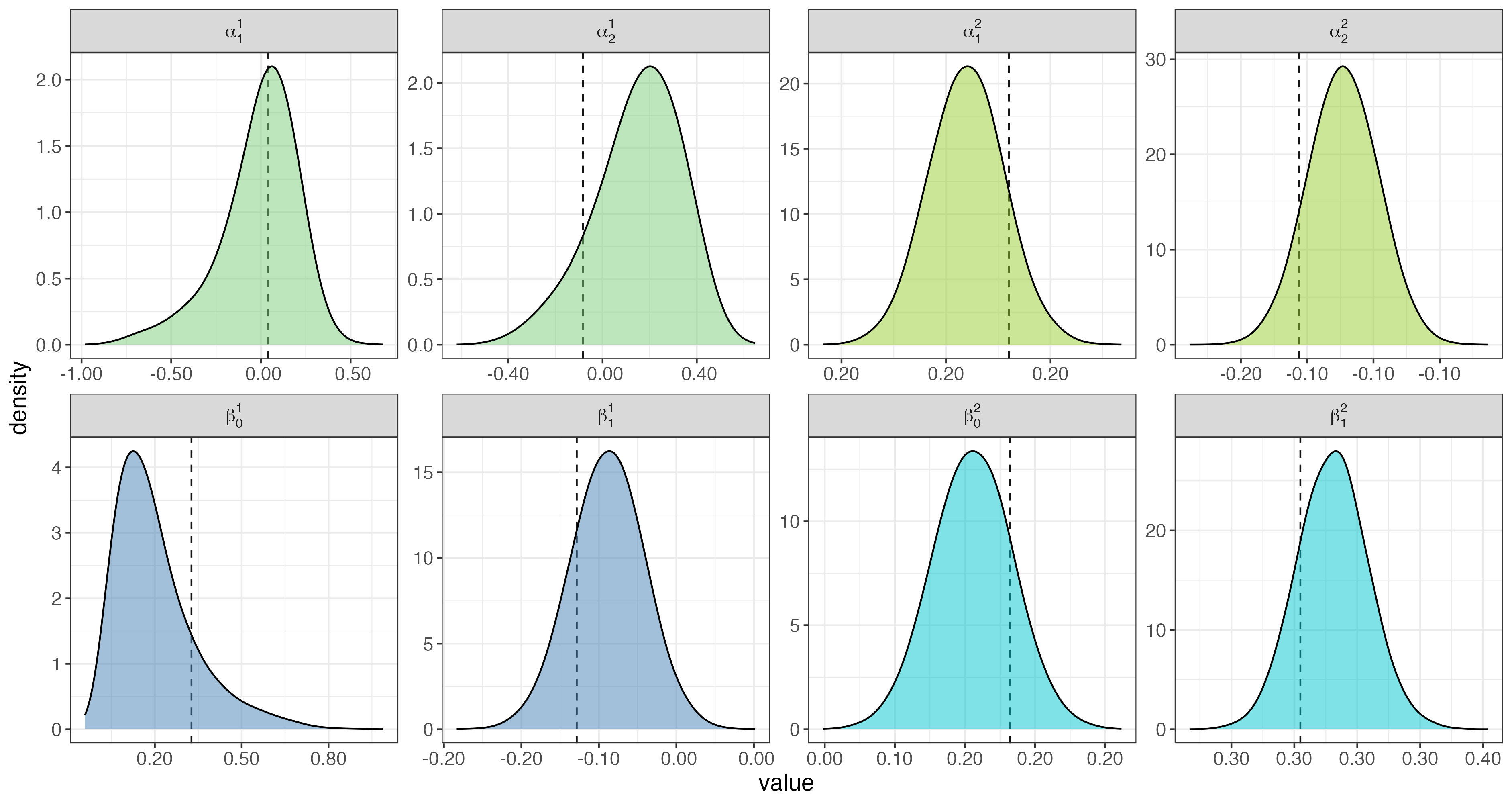}
     \includegraphics[width = 0.8\linewidth]{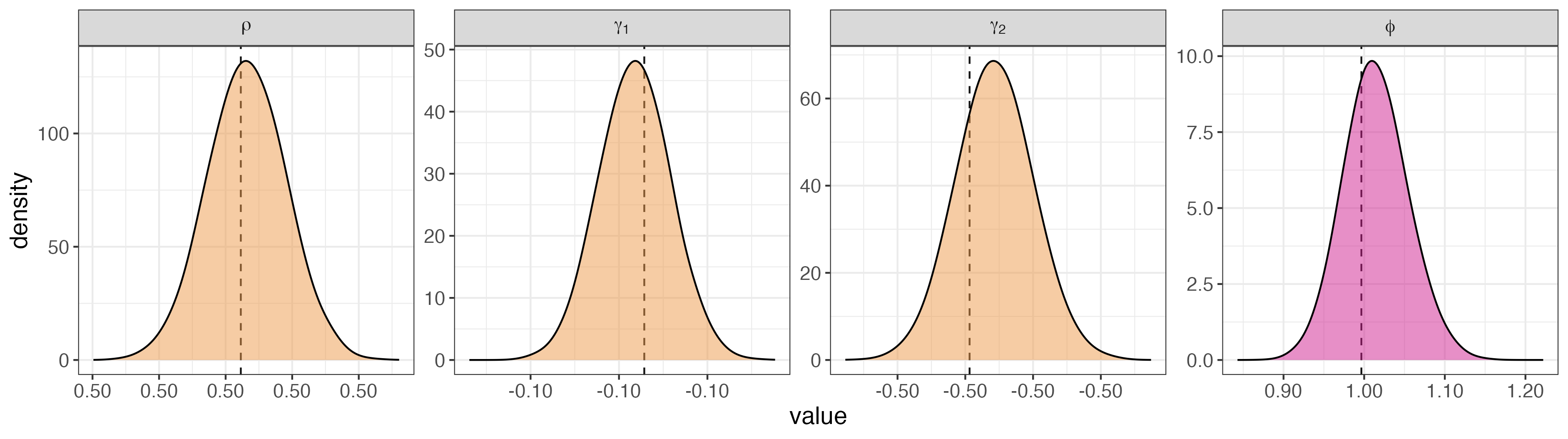}
    \caption{Posterior distribution of the Poisson-Gamma Granger GLM model parameters fitted to an artificial bivariate trajectory of length $n=1000$. The data is simulated using the parameter values indicated with vertical dashed lines, and single-site MH algorithms are used to sample $p(\theta_{i}|\boldsymbol{\theta}_{-i}, \boldsymbol{y_1}, \boldsymbol{y_2})$ for all $i$.}
    \label{fig:bayes_sim1}
\end{figure}

We finish off this section with a simulated example that illustrates the Bayesian fit of a Granger-GLM model using the algorithms just described. To this end, a bivariate causal trajectory of length $n=1000$ is drawn from the Poisson-Gamma model with $T_1(y) = \log y$ and $T_2(y) = \log(y +1)$ and parameter values $\boldsymbol\beta^{(1)} = (0.1, -0.1)^\top, \boldsymbol\beta^{(2)} = (0.2, 0.3)^\top, \boldsymbol\alpha^{(1)} = (0.2, 0.4)^\top, \boldsymbol\alpha^{(2)}= (0.2, -0.1)^\top, \boldsymbol\gamma = (-0.1, -0.5)^\top,\phi_1 = 1$ and $\rho = 0.5$. Prior variances are set to $100$ for all $\boldsymbol{\theta}$ elements, and a vanishing calibration of the proposals scaling values $\boldsymbol{\sigma}_{1}, \boldsymbol{\sigma}_{1}$ is taken, targeting about 44\% acceptance rate of each element. Running the Markov Chain for a total of 11K iterations and discarding the first 1K as burn-in period renders the samples used to estimate the posterior distributions in Figure \ref{fig:bayes_sim1}. The parameter values used to simulate the data as shown as vertical dashed lines, all within regions of high posterior probability, as expected. 

Given the successful application of standard MCMC techniques to infer the posterior bivariate Granger-GLM models, we extend the methodology to the causality selection task in the upcoming Subsection \ref{sec:spike_slab}. 


\section{Testing for Granger causality}\label{sec:GC}

In this section, we develop a procedure for testing if $\{Y_{1\,t}\}_{t\in\mathbb N}$ Granger causes $\{Y_{2\,t}\}_{t\in\mathbb N}$. 
In terms of formal statements of the hypotheses: $\texttt H_0: \boldsymbol\gamma={\bf0}\,\,\mbox{and}\,\,\rho=0$ ($\{Y_{1\,t}\}_{t\in\mathbb N}$ does not Granger cause $\{Y_{2\,t}\}_{t\in\mathbb N}$) against $\texttt H_1: \gamma_{l} \neq0\,\,\mbox{for some}\,\,l=1,\ldots,k\,\,\mbox{or}\,\,\rho\neq0$ (there is Granger causality). We briefly discuss the frequentist case based on the likelihood ratio test and then focus our attention on the Bayesian framework including the important task of causal order selection.

\subsection{Likelihood ratio test}

We develop a procedure for testing the above hypothesis using the likelihood ratio test. Consider the parametric time series GLM, define $\boldsymbol\theta_0=(\boldsymbol\beta^{(1)}, \boldsymbol\beta^{(2)}, \boldsymbol\alpha^{(1)}, \boldsymbol\alpha^{(2)}, {\bf0}_k,\phi_1,\phi_2,0)^\top$, with $\widehat{\boldsymbol\theta}_0$ being its corresponding maximum likelihood estimator.
The likelihood ratio statistic for testing Granger causality is given by $\mbox{LR}=2\left(\ell(\widehat{\boldsymbol\theta})-\ell(\widehat{\boldsymbol\theta}_0)\right)$.
Under certain regularity conditions, including that which $\texttt H_0$ does not belong to the boundary of the parameter space, we have that $\mbox{LR}$ are asymptotically $\chi^2_{k+1}$ distributed (chi-square with $k+1$ degrees of freedom) under the null hypothesis.

We perform a Monte Carlo simulation to evaluate the performance of the proposed LR test for Granger causality. We generate 1000 replications from the Poisson-gamma time series model under the configuration given in the previous section but with $\gamma_1=\gamma_2=\rho=0$, that means that there is no Granger-causality. Table \ref{tab:confI_siglevel} gives us the empirical significance levels of the likelihood ratio test for $n=500,1000,2000$ and under nominal significance levels at 1\%, 5\%, and 10\%. The simulated results show that the empirical significance levels are approaching the nominal ones as the sample size increases. We can also observe that the LR test does not produce satisfactory results for $n=500$. The results for $n=1000$ and $n=2000$ are satisfactory. Therefore, conclusions on Granger causality based on LR tests should be taken carefully for small sample sizes.

\begin{table}[h!]
\centering
\begin{tabular}{cccc}
\hline
$n\rightarrow$ & 500   & 1000  & 2000  \\
\hline
1\%            & 0.019 & 0.015 & 0.009 \\
5\%            & 0.096 & 0.062 & 0.058 \\
10\%           & 0.156 & 0.109 & 0.121\\
\hline
\end{tabular}\caption{Empirical significance levels produced by the likelihood ratio test for checking Granger causality when absent. The sample sizes are $n=500,1000,2000$.}\label{tab:confI_siglevel}
\end{table}


\subsection{Bayesian approach with spike and slab priors}\label{sec:spike_slab}

In this section, we explore the idea of using spike and slab priors to carry out dependency order selection (lag $k$) for Granger-GLMs. Spike and slab are sparsity-inducing priors, often applied in the context of regression to decide the predictors to be included in the model. See for instance \cite{mitchelbeauchamp1988} and \cite{rao2005}.

Spike and slabs are hierarchical priors that enable a fully Bayesian approach to model selection. In this approach, the probability of inclusion of candidate covariates is properly quantified. One type of spike and slab priors known as \textit{Dirac spikes} is constructed as a two-point mixture. The \textit{spike} piece places point mass at zero,  encouraging variable exclusion. The \textit{slab} component is some diffuse distribution that models the prior probability of the variables included in the model. Another popular possibility referred to as \textit{continuous spikes} sets unimodal prior distributions with mode zero for both spike and slab components. However, the slab-to-spike variance ratio is usually considerably lower than one. 

The covariates inclusion or exclusion is modeled with a set of indicators $\boldsymbol{\delta} = (\delta_1, \delta_2, \ldots )$, where $\delta_i \in \{0, 1\}$ and predictor $i$ is in the slab component if $\delta_i =0$. Otherwise if $\delta_i =0$, $i$ is allocated to the spike. In what follows, a spike and slab prior for $\boldsymbol{\gamma}$ is formulated and inference is outlined in detail. By doing this, our goal is to perform the selection of $k$, the order of the lagged causal time series ${Y}_{1, t-1}, {Y}_{1, t-2}, \ldots, {Y}_{1, t-k}$ in the conditional mean of $Y_{2,t}$. In other words, a maximum $k \equiv k_{max}$ is chosen, and the model is fitted with given $(p,q,r,s)$ and $k_{max}$. Sparsity inducing priors for $\gamma_1, \ldots, \gamma_{k_{max}}$ will then shrink towards zero the effects of $Y_{1, t-j}$ that are not relevant for $\nu_{2,t}$.  

Spike and slab priors are easily accommodated within MCMC by augmenting the posterior with the unobserved vector $\boldsymbol{\delta}$, here $\boldsymbol{\delta} = (\delta_1, \ldots, \delta_{k_{max}})^\top$. First, the prior probability of inclusion, $P(\delta_i=1|\omega)$, is Bernoulli($\omega$) for all $i$, where uncertainty on $\omega$ is accommodated with a hyperprior specification, $\omega \sim \mbox{Beta}(a,b)$, $a,b>0$. The hyperparameters $a,b$ are pre-specified and $a = b = 1$ render $\omega \sim \mbox{Uniform}(0,1)$.

The spike and slab prior for $\boldsymbol{\gamma}$ is conditional on $\boldsymbol{\delta}$ and can be written as
\begin{eqnarray}
    p( \boldsymbol{\gamma}| \boldsymbol{\delta}) = \prod_{i=1}^{k_{max}} p( {\gamma_i}| {\delta}_i) = \prod_{i=1}^{k_{max}}\left\{ \delta_i\times \pi_{slab}(\gamma_i) +  \delta_0(\gamma_i)\right\},
\end{eqnarray}
where $\delta_0(\gamma_i)$ is the Dirac function.

To incorporate the spike and slab approach in the algorithms developed in Section \ref{sec:bayes}, consider the augmented posterior of $\boldsymbol{\theta}_2$ given by
\begin{equation}
    p(\boldsymbol{\theta}_2, \boldsymbol{\delta}, \omega | \boldsymbol{\theta}_1, \boldsymbol{y}_1, \boldsymbol{y}_2) \propto f(\boldsymbol{y}_1, \boldsymbol{y}_2 |\boldsymbol{\theta}_1, \boldsymbol{\theta}_2) \pi(\boldsymbol{\theta}_{2, -\boldsymbol{\gamma}}) p(\boldsymbol{\gamma}| \boldsymbol{\delta})  p(\boldsymbol{\delta}| \boldsymbol{\omega}) p(\omega)   .
\end{equation}

Sampling from (\ref{eq:mh_y2}) is preceded by drawing from $p(\omega|\ldots)$ and $p(\boldsymbol{\delta}|\ldots)$ which are the full-conditional distributions of $\omega$ and $\boldsymbol{\delta}$. The full-conditional of $\omega$ is

\begin{eqnarray}
    p(\omega| \ldots ) \propto p(\boldsymbol{\delta}|\omega)p(\omega) = \omega^{\sum_{i=1}^{k_{max}} \delta_j}(1-\omega)^{k_{max} - \sum_{i=1}^{k_{max}} \delta_j} \omega^{a-1} (1-\omega)^{b-1},
\end{eqnarray}
which we recognize as the kernel of a Beta distribution with shape $a + \sum_{i=1}^{k_{max}} \delta_j$ and rate $b + \sum_{i=1}^{k_{max}} \delta_j + k_{max}$. Accordingly, $\omega$ can be simulated given the current $\boldsymbol{\delta}$ with a Gibbs update, i.e., $\omega^{t+1} \sim$ Beta($a + \sum_{i=1}^{k_{max}} \delta_j^t$, $b + \sum_{i=1}^{k_{max}} \delta_j^t + k_{max}$). 

Under the current states of $\omega, \boldsymbol{\theta}_1$ and $\boldsymbol{\theta}_2$, the update of $\boldsymbol{\delta}$ consists of sampling from the full-conditional distribution $p(\boldsymbol{\delta}| \ldots) \propto f(\boldsymbol{y}_1, \boldsymbol{y}_2|\boldsymbol{\theta}_1, \boldsymbol{\theta}_{2,\boldsymbol{\delta}}) p(\boldsymbol{\delta}|\omega)$. The notation $\boldsymbol{\theta}_{2,\boldsymbol{\delta}}$ is used to indicate that inclusion of $\gamma_i$ is subject to $\delta_i = 1$, or equivalently that $\nu_{2\,t}$ is

$$\nu_{2\,t} = \beta_0^{(2)}+\sum_{i=1}^p \beta_i^{(2)}\widetilde Y_{2\, t-i}+\sum_{j=1}^q \alpha_j^{(2)}\nu_{2\, t-j}+\sum_{l=1}^k \gamma_l  \delta_l  \widetilde Y_{1\, t-l}.$$

Sampling $\boldsymbol{\delta}$ can be done via a full-sweep Gibbs algorithm that draws sequentially from $p(\delta_i|\boldsymbol{\delta}_{-i}, \ldots)$, where $\boldsymbol{\delta}_{-i}$ denotes $\boldsymbol{\delta}$ without element $i$, for $i = 1, \ldots, k_{max}$. This follows from obtaining the probability of inclusion $p(\delta_i=1| \boldsymbol{\delta}_{-i}, \ldots)$ from the normalisation 

\begin{eqnarray}\label{eq:prob_delta}
p(\delta_i =1| \boldsymbol{\delta}_{-i}, \ldots) = \frac{p(\delta_i=1| \boldsymbol{\delta}_{-i}, \ldots)}{\sum_{j=0}^1 p(\delta_i=j| \boldsymbol{\delta}_{-i}, \ldots)} = \frac{ \omega f(\boldsymbol{y}_1, \boldsymbol{y}_2|\boldsymbol{\theta}_1, \boldsymbol{\theta}_{2,\boldsymbol{\delta^1}}) }{ \omega f(\boldsymbol{y}_1, \boldsymbol{y}_2|\boldsymbol{\theta}_1, \boldsymbol{\theta}_{2,\boldsymbol{\delta^1}}) +  (1-\omega) f(\boldsymbol{y}_1, \boldsymbol{y}_2|\boldsymbol{\theta}_1, \boldsymbol{\theta}_{2,\boldsymbol{\delta^0}})}
\end{eqnarray}
where $\boldsymbol{\delta^1}$ and $\boldsymbol{\delta^0}$ denote the current $\boldsymbol{\delta}$ vector with element $i$ set as $\delta_i =1$ or $\delta_i =0$, respectively. Equation (\ref{eq:prob_delta}) is then the probability of including $Y_{1, t-i}$ in $\nu_{2,t}$, which is a weighted ratio of the data likelihood with inclusion or exclusion of $Y_{1, t-i}$. A full-sweep update renders $\boldsymbol{\delta}^{t+1}$ by sampling $\delta^{t+1}_i$ sequentially for $i=1,\ldots, k_{max}$ as Bernoulli trials with success probability (\ref{eq:prob_delta}). 

Given delta, the caused time series parameters $\boldsymbol{\theta}_2$ are sampled from the full-conditional  $p(\boldsymbol{\theta}_2, | \boldsymbol{\delta}, \boldsymbol{\theta}_1, \boldsymbol{y}_1, \boldsymbol{y}_2) \propto  f(\boldsymbol{y}_1, \boldsymbol{y}_2 |\boldsymbol{\theta}_1, \boldsymbol{\theta}_2, \boldsymbol{\delta}) \pi(\boldsymbol{\theta}_{2, -\boldsymbol{\gamma}}) p(\boldsymbol{\gamma}| \boldsymbol{\delta})  p(\boldsymbol{\delta}| \boldsymbol{\omega})$ which is amenable to the sampling algorithms of Section \ref{sec:bayes}. The term $\pi(\boldsymbol{\theta}_{2, -\boldsymbol{\gamma}})$ denotes the prior distribution of $\boldsymbol{\theta}_2$ parameters different than $\boldsymbol{\gamma}$, i.e, $\boldsymbol{\beta}^{(2)}, \boldsymbol{\alpha}^{(2)}, \rho, \phi_2$. When $k \in \{p+q+2, \ldots, p+q+1+k_{max}\}$ and the update is of $(\gamma_1, \ldots, \gamma_{k_{max}})$, only $\gamma_i$ allocated to the slab component at the current iteration are sampled once $p(\gamma_i|\delta_1 = 0) = 0$.

The strategies just outlined to sample the full-conditional distributions of a causal Granger-GLM model with spike and slab priors for $\boldsymbol{\gamma}$ are summarized with pseudocode in Algorithm \ref{alg:summary_algorithm}. The algorithm's inputs are the observed data ($\boldsymbol{y}_1, \boldsymbol{y}_2$), proposal variances of $\boldsymbol{\theta}_1$ and $\boldsymbol{\theta}_{2, -\boldsymbol{\gamma}}$ ($\boldsymbol{\sigma}_1,  \boldsymbol{\sigma}_2$), prior hyperparameters ($\boldsymbol{\tau}, a ,b$), and orders $(p,q,r,s,k_{max})$.

\begin{algorithm}
\DontPrintSemicolon
\SetAlgoLined
\SetNoFillComment
\LinesNotNumbered 
\KwIn{Observed data $\boldsymbol y_1, \boldsymbol{y}_2$; proposal scaling $\boldsymbol{\sigma}_1,  \boldsymbol{\sigma}_2$; hyperparameters $\boldsymbol{\tau}^2,a,b$; orders $(p,q,r,s,k_{max})$; current states $\boldsymbol{\theta}^t, \omega^t, \boldsymbol{\delta}^t$;}

\vspace{0.3cm}

\tcp*[l]{1. ${\boldsymbol{\theta}_1}$ MH-update}
\texttt{\textcolor{cyan}{ \footnotesize Samples $\boldsymbol{\theta}^1 \equiv ( \boldsymbol{\beta}^{(1)}, \boldsymbol{\alpha}^{(1)}, \phi_1) $}}

\For{$j \leftarrow 1:(p +q +2)$}{ 
    
    Draw $\theta_{1, j}' \sim q_{1,j}(\cdot| \theta^t_{1, j} )$ and set $\boldsymbol{\theta_1}' \equiv \boldsymbol{\theta}^{t}[j] \leftarrow \theta_{1, j}'$;

    Compute $ p(\theta_{1,j}'|\theta_{1,j}^t) = \min \left\{1, \frac{f(\boldsymbol{y}_1|\boldsymbol{\theta}_1') \pi(\theta_{1,j}') q_j( \theta_{1,j}^t| \theta_{1,j}') }{f(\boldsymbol{y}_1|\boldsymbol{\theta}_1^t) \pi(\theta_{1,j}^t) q_j( \theta_{1,j}'| \theta^t_{1,j})} \right\}$;

    With probability $p(\theta_{1,j}'|\theta_{1,j}^t)$, set ${\theta}_{1,j}^{t} \leftarrow {\theta}_{1,j}'$, otherwise ${\theta}_{1,j}^{t} \leftarrow {\theta}_{1,j}^t$; 

}

\vspace{0.3cm}

\tcp*[l]{2. $\omega$ Gibbs update}
Sample $\omega^{t} \sim \mbox{Beta}(a + \sum_{i=1}^{k_{max}} \delta_j^t$, $b + \sum_{i=1}^{k_{max}} \delta_j^t + k_{max}$);

\vspace{0.3cm}

\tcp*[l]{3. ${\boldsymbol{\delta}}$ Gibbs update}

\For{$i \leftarrow 1:k_{max}$}{ 
Compute the inclusion probability $p_{\delta^t_i} \equiv p(\delta^t_i = 1|\boldsymbol{\delta}^t_{-i}, \boldsymbol{\theta}_1^{t}, \boldsymbol{\theta}_2^t, \omega^t)$ using Equation (\ref{eq:prob_delta});

Update $\delta^t_i$ by drawing from a Bernoulli with success probability $p_{\delta^t_i}$;

}

\vspace{0.3cm}

\tcp*[l]{4. ${\boldsymbol{\theta}_2}$ MH-update}

\vspace{0.15cm}

\For{$j  \leftarrow 1:p+q+k_{max}+2$}{ 

\texttt{\textcolor{cyan}{\footnotesize $\boldsymbol{\gamma}$ elements for included lags: }}

  \uIf{ $j \in (p+q+2):(p+q+1+k_{max}$)}{ 

  \uIf{$\delta^{t} =1$}{

    Draw $\theta_{2, j}' \sim q_{2,j}(\cdot| \theta^t_{2, j} )$;
    
    Set $\boldsymbol{\theta_2}' \equiv \boldsymbol{\theta}_2^{t}[j] \leftarrow \theta_{2, j}'$;

    Compute $p(\theta_{2,j}'|\theta_{2,j}^t) = \min \left\{1, \frac{f(\boldsymbol{y}_1, \boldsymbol{y}_2 |\boldsymbol{\theta}_1^{t+1}, \boldsymbol{\theta}_2', \boldsymbol{\delta}^{t+1}) \pi_{slab}(\theta_{2,k}')}{f(\boldsymbol{y}_1, \boldsymbol{y}_2|\boldsymbol{\theta}_1^{t}, \boldsymbol{\theta}_2^t, \boldsymbol{\delta}^{t}) \pi_{slab}(\theta_{2,k}^t)} \right\}$ and set ${\theta}_{2,j}^{t} \leftarrow {\theta}_{2,j}'$ with probability $p(\theta_{2,j}'|\theta_{2,j}^t)$, otherwise ${\theta}_{2,j}^{t} \leftarrow {\theta}_{2,j}^t$; 
  
  } 
  \Else{
  \texttt{skip};
  }

\vspace{0.15cm}
    
  }
  \texttt{\textcolor{cyan}{\footnotesize Update of $\boldsymbol{\theta_{2, -\boldsymbol{\gamma}}} \equiv ( \boldsymbol{\beta}^{(2)}, \boldsymbol{\alpha}^{(2)}, \rho, \phi_2)$: }}
  
  \Else{
    
    Propose $\theta_{2, j}'$ by sampling from $q_{2,j}(\cdot| \theta^t_{2, j} )$ and set $\boldsymbol{\theta_2}' \equiv \boldsymbol{\theta}_2^{t}[j] \leftarrow \theta_{2, j}'$

    Compute $p(\theta_{2,k}'|\theta_{2,k}^t) = \min \left\{1, \frac{f(\boldsymbol{y}_1, \boldsymbol{y}_2 |\boldsymbol{\theta}_1^{t}, \boldsymbol{\theta}_2', \boldsymbol{\delta}^{t}) \pi(\theta_{2,k}') q_k( \theta_{2,k}^t| \theta_{2,k}') }{f(\boldsymbol{y}_1, \boldsymbol{y}_2|\boldsymbol{\theta}_1^{t}, \boldsymbol{\theta}_2^t, \boldsymbol{\delta}^t) \pi(\theta_{2,k}^t) q_k( \theta_{2,k}'| \theta^t_{2,j})} \right\}$ and set ${\theta}_{2,j}^{t} \leftarrow {\theta}_{2,j}'$ with probability $p(\theta_{2,j}'|\theta_{2,j}^t)$, otherwise ${\theta}_{2,j}^{t} \leftarrow {\theta}_{2,j}^t$;

  }

}

\caption{Sampling the posterior distribution of a bivariate causal Granger-GLM with Dirac spike and slab priors for $\boldsymbol{\gamma}$ to select causality order. }\label{alg:summary_algorithm}
\end{algorithm}

\begin{figure}[h!]
    \centering
    \includegraphics[width = 0.6\linewidth]{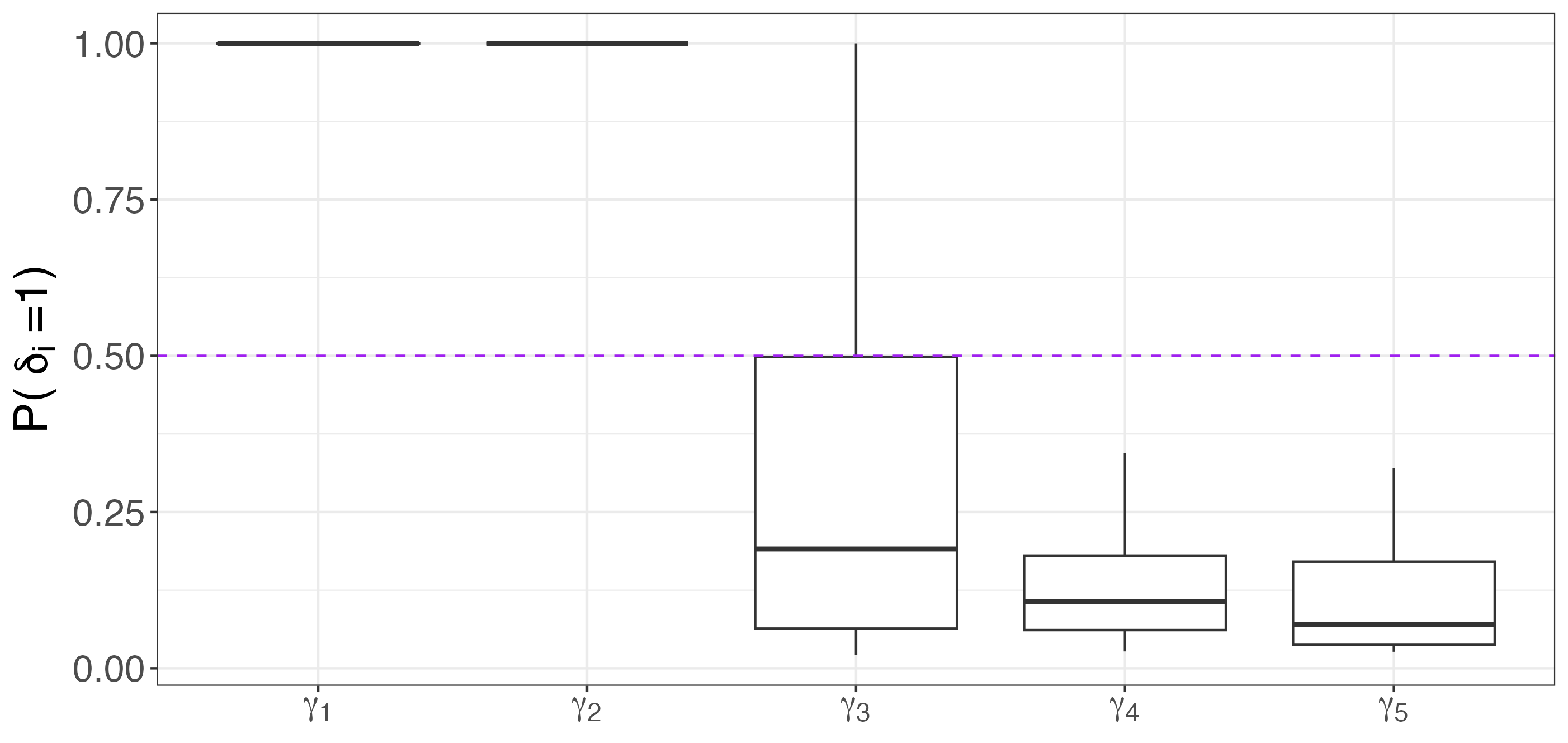}
    \caption{Boxplots of the inclusion probabilities $P(\delta_i = 1)$ for $i = 1, \ldots, 5$ obtained from one hundred simulated Poisson-Gamma data sets with $\boldsymbol{\gamma} = (-0.1,-0.5)^\top$. }\label{fig:spike_slab_sim1}
\end{figure}

Simulated experiments are conducted to assess the performance of the spike and slab selection of $\boldsymbol{\gamma}$ elements for known $k$ and $k_{max} > k$. Using the parameter configuration of the Poisson-Gamma model of Section \ref{sec:bayes}, we simulate 100 Granger-causal data sets that are fitted with Algorithm \ref{alg:summary_algorithm} and $k_{max} =5$. In this example, $\boldsymbol{\gamma} = (-0.1,-0.5)$ so the true $k$ is 2. For each simulated data, we discard the first 1K iterations as a burn-in period and estimate inclusion probabilities $P(\delta_i = 1)$ for $i=1, \ldots, 5$ with Monte Carlo averages. Figure \ref{fig:spike_slab_sim1} summarises the results with boxplots of the one hundred $P(\delta_i = 1)$ values, with a horizontal dashed line shown at probability 0.5. Inclusion probabilities for orders one and two are all near one, indicating that $Y_{1, t-1}$ and $Y_{1, t-2}$ should be included as covariates that explain the mean $Y_{2,t}$. Consistently with the configuration used to simulate the data, $P(\delta_i = 1)$ is smaller for $i=3$, decreasing substantially when $i=4,5$. 

\begin{figure}[h!]
    \centering
    \includegraphics[width = 0.6\linewidth]{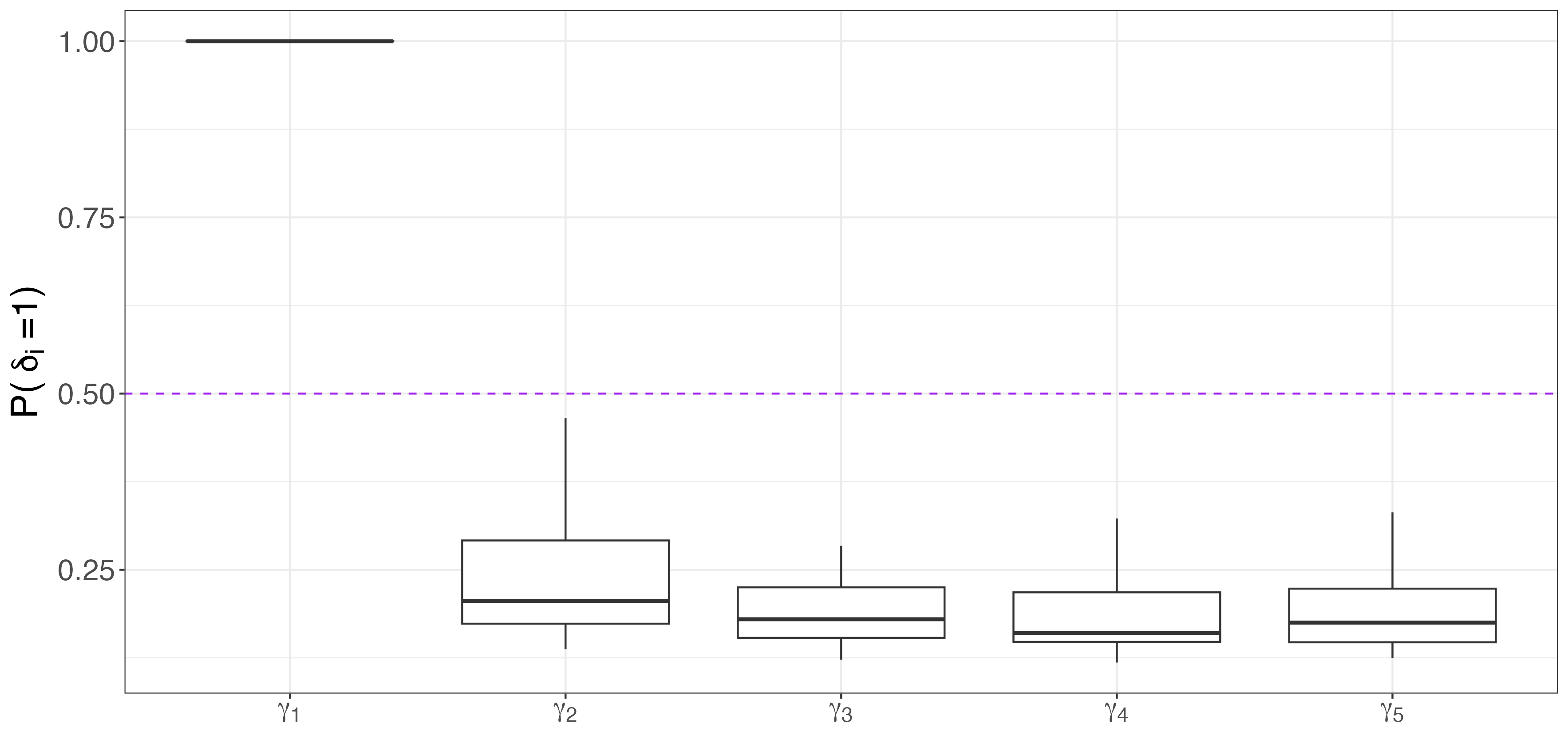}
    \caption{Boxplots of the inclusion probabilities $P(\delta_i = 1)$ for $i = 1, \ldots, 5$ obtained from one hundred simulated Geometric-Geometric data sets with $\boldsymbol{\gamma} = -0.5$. } \label{fig:spike_slab_sim}
\end{figure}

A second simulation experiment is designed with a Geometric-Geometric model and $\boldsymbol{\gamma} = -0.5$. Results illustrated in Figure \ref{fig:spike_slab_sim} show an excellent performance of the spike and slab approach, which indicates $k=1$, as expected. In the following section, we apply the Bayesian inferential approaches developed and examined in Sections \ref{sec:bayes} and \ref{sec:spike_slab} to actual bivariate time series derived from brain recordings.

\begin{remark}
In a similar manner, it is possible to formulate a spike and slab approach for order selection of $\boldsymbol{\beta^{(1)}}$ ($p$), $\boldsymbol{\alpha}^{(1)}$ ($q$), $\boldsymbol{\beta}^{(2)}$ ($r$) and $\boldsymbol{\alpha}^{(2)}$ ($s$). Although it works very well for $p$ and $r$ as with $k$, it is problematic when considering $q$ and $s$. We have found from experiments that inclusion probabilities for $\boldsymbol{\alpha}^{(1)}$ and $\boldsymbol{\alpha}^{(2)}$ elements used to simulate the data are lower than expected. Our conjecture is that this happens because $\boldsymbol{\alpha}^{(1)}$ and $\boldsymbol{\alpha}^{(2)}$ are coefficients associated with latent variables, which induces further uncertainty. On the other hand, $\boldsymbol{\beta}^{(1)}$, $\boldsymbol{\beta}^{(2)}$ and $\boldsymbol{\gamma}$ are effects associated with observed data, which facilitates the correct order identification. For this reason, we focused the order selection task on the causal component that is of main interest.
\end{remark}


\section{Causality in Brain Activity}\label{sec:braindataapp}

In this section, we present the Granger-causality studies on the brain activity data described in Subsection \ref{intro:data}. Spectral power (derived from periodograms), which are strictly positive-valued, is jointly analyzed with the number of spikes using the Granger-causal Poisson-Gamma model. Results obtained with the proposed model and the spike and slab approach for inference are reported in Section \ref{sec:poisson_gamma}. The Poisson-Gamma fit to the transformed data is explored and interpreted in detail. In Subsection \ref{sec:geo_geo}, the pairwise dependencies between spike trajectories from five different electrodes are investigated via a Granger-causal geometric-geometric model.


\subsection{LFP power at $\beta$ bands and spike counts data analysis}\label{sec:poisson_gamma}

Granger-causality between the LFP power at the $\beta$ band and spike counts in the same interval is now investigated in the LFP $\rightarrow$ spike direction. We set $\{Y_{1t}\}_{t=1}^{133}$ to be the marginal LFP $\beta$-power trajectory, which Granger-causes spike counts, $\{Y_{2t}\}_{t=1}^{133}$. The Gamma-Poisson case models the above continuous positive and integer-valued time series as 
\begin{equation}\label{eq:poisson_gamma}
   \begin{cases}
            Y_{2t}|\mathcal{F}^{(1,2)}_{t-1} \sim \mbox{Poisson}( \mu_{2t} \exp(\rho Y_{1t})),\\
         \mu_{2t} = \exp\left( \beta_0^{(2)} + \sum_{i=1}^p \beta_i^{(2)}\log(Y_{2\,t-i} +1) + \sum_{j=1}^q \alpha_j^{(2)} \log(\mu_{2\,t-1}) + \sum_{l=1}^k \gamma_l \log(Y_{1\,t-l}) \right), \\
            Y_{1t}|\mathcal{F}^{(1)}_{t-1} \sim \mbox{Gamma}( \mu_{1t}, \phi_1), \\
         \mu_{1t} =   \exp\left( \beta_0^{(1)} + \sum_{i=1}^r\beta_i^{(1)}  \log(Y_{1\,t-i} + 1) + \sum_{j=1}^s \alpha_j^{(1)} \log(\mu_{1\,t-1}) \right), \\
    \end{cases}
\end{equation}
where the gamma distribution takes the GLM parameterization. The original and processed data as described previously are displayed in Figure \ref{fig:lfp_spike_fig}. 

Model (\ref{eq:poisson_gamma}) is fitted to the processed data using the Bayesian spike-slab approach with $p=q=r=s=1$ and $k_{max}=15$. The posterior distribution of model parameters and latent variables is estimated with 100K iterations of Algorithm \ref{alg:summary_algorithm}. With posterior samples of $\delta_k$, we can estimate the probability that the beta spectral power at lags $1,\ldots, 15$ time units, i.e. $Y_{2,t-1}, \ldots, Y_{2, t-15}$, are causal of the spiking rate $\mu_{1t}$ via Monte Carlo averaging. Given that one-time unit corresponds to 30 milliseconds, the configuration $k_{max} = 15$ scans the LFP power from 30 to 450 ms before the spiking rate. Our main aim is to identify causality lag(s) that have a high probability of causing $\mu_{2t}$, a task well facilitated by the spike and slab approach. This is quantified via $P(\delta_k =1)$ which is estimated by $\sum_{i=1}^{100K} \delta_k^i/100K$ where $\delta_k^1, \ldots, \delta^{100K}_k$ are $\delta_k$ posterior samples. 

In addition, we would like to compare the causality selection done with our model to other standard statistical approaches. To that end, four alternative models will be considered. The first is a special case of the bivariate Poisson-Gamma model which is obtained under the assumption that the causal series $Y_1$ is observed. Denoted by \textit{Poisson Spike Slab (SS)}, this is a univariate Poisson regression with structure (\ref{eq:poisson_gamma}) (for $Y_2$) that also employs a spike and slab (SS) prior for $\boldsymbol{\gamma}$. The other two models included in our analysis are based on the integer-valued generalized autoregressive conditional heteroscedasticity (INGARCH) approach. The INGARCH class is a well-cemented tool for count time series data that assume the general form $g(\mu_t) = \beta_0 + \sum_{k=1}^p \beta_k \widetilde{g}(Y_{t-k}) + \sum_{l=1}^q g(\mu_{t-l}) + \boldsymbol{\gamma}^T\boldsymbol{X}_t$, where  $g(\cdot)$ is a link function, $\tilde{g}(\cdot)$ is some transformation and $\mu_t = E(Y_t|\mathcal{F}_{t-1})$. Popular choices for the conditional distribution $Y_t|\mathcal{F}_{t-1}$ are the Poisson and Negative Binomial. We consider these two with $p=q=1$, $g(x) \equiv \log(x)$, $\tilde{g}(x)\equiv \log(x +1)$ and regression structure ($\boldsymbol{\gamma}^T\boldsymbol{X}_t$) given by $\gamma_1 Y_{1, t-1} + \ldots + \gamma_1 Y_{t-15}$. In other words, the lagged LFP effects are included as observed covariates, as done for the \textit{Poisson (SS)} case. The Poisson-INGARCH(1,1) and Negative-Binomial INGARCH(1,1) models will be fitted using the \texttt{R} package \texttt{tscount}. Finally, the last model is an AutoRegressive Moving Average with covariates, ARMAX(1,1). This is similar in structure to the INGARCH models but is based on a Gaussian assumption for $Y_t|\mathcal{F}_{t-1}$. Naturally, this is the poorest choice amongst the models just described because the response variable (spikes) assumes low counts. 

\begin{figure}
    \centering
    \includegraphics[width=0.85\linewidth]{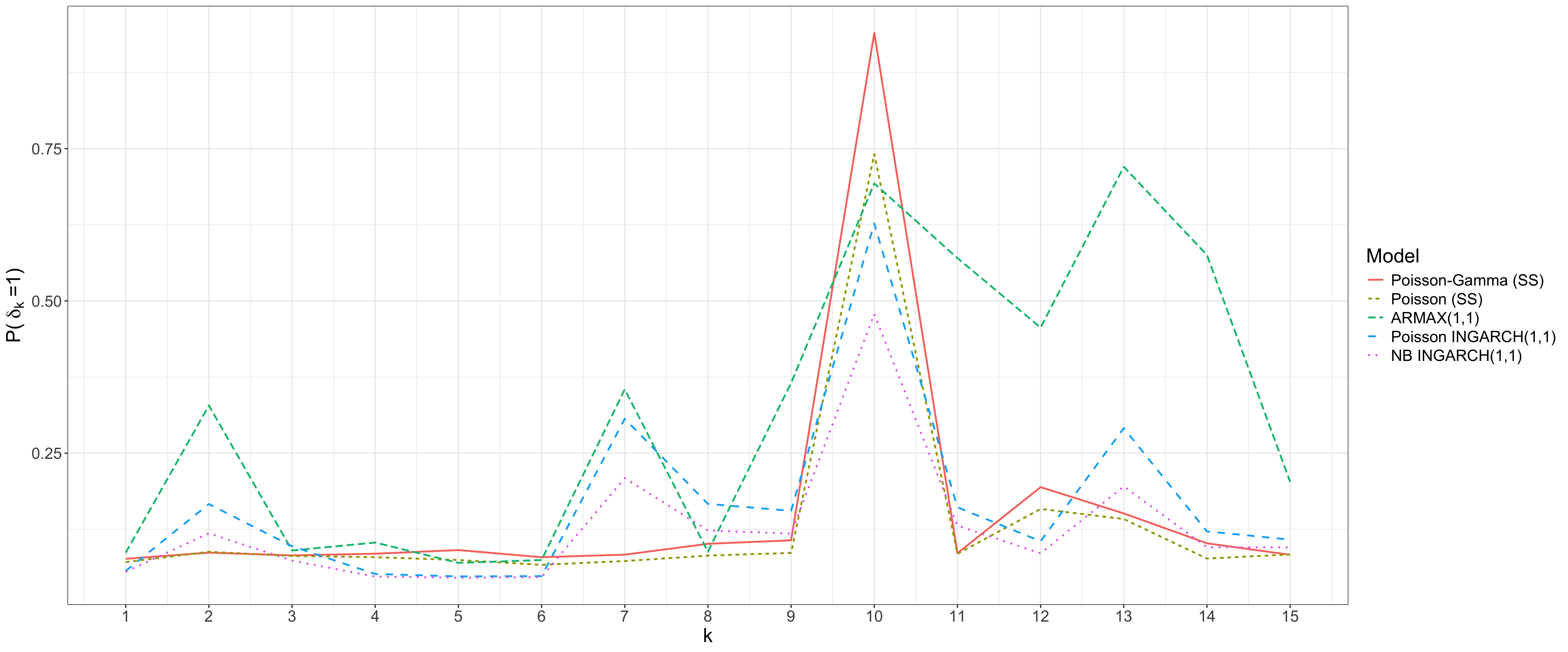}
    \caption{Estimated probability of non-zero effects of $Y_{2, t-1}, \ldots, Y_{2, t-15}$ in the time $t$ spiking rate $\mu_{1t}$ obtained under different models. }
    \label{fig:delta_selection}
\end{figure}

Figure \ref{fig:delta_selection} displays the results from the different models, with $k=1, \ldots, 15$ on the horizontal axis and the estimated probability $P(\delta_k =1)$ on the vertical axis. Different line types are used to indicate the five approaches employed for the causality selection. One remark here is that $P(\delta_k =1)$ is immediately available only from models fitted with the Bayesian spike and slab approach (\textit{Poisson-Gamma (SS)} and \textit{Poisson (SS)}) which is in fact one methodological contribution from this paper. In order to give estimated inclusion probabilities for the INGARCH and ARMAX, the following strategy is employed. First, the models are fitted to the data using the available implementations in \texttt{R} that rely on the frequentist approach. To estimate inclusion probabilities for $\gamma_1, \ldots, \gamma_{15}$ a parametric bootstrap procedure is implemented as follows. First, a trajectory is simulated from the fitted model, and its parameters are estimated. Then, 95\% confidence bands are created for these estimates, and the inclusion of $\gamma_l$  is counted for that replica if and only if its 95\% confidence interval does not contain zero.  This is repeated 10K times for each fitted model (\textit{ARMA(1,1), Poisson INGARCH(1,1), NB INGARCH(1,1)}) and the percentage of times that each $\gamma_1, \ldots, \gamma_{15}$ was non-zero is reported. 

In Figure \ref{fig:delta_selection}, the \textit{Poisson-Gamma (SS)} the model provides evidence that the beta spectral power (at lag 10-time units), denoted $Y_{1, k-10}$, is causal
of the spiking rate $\mu_{2t}$ with a high probability of 94\%. This indicates that activity in the $\beta$ band is predictive at about 300 milliseconds before the spike. INGARCH-based models are in accordance in the sense that their highest inclusion probabilities are also at $k=10$. In contrast to other models, evidently, there is less certainty of this under the \textit{Poisson-Gamma (SS)}. Results from the ARMAX(1,1) model are less parsimonious and indicate $k=10$ and $k=13$, with similar probability. This poses the question of whether it is adequate to consider only $k=10$ or both $k=10, 13$. We evaluate this by looking at information criteria (AIC and BIC) which in fact indicates that the ARMAX(1,1) provides the worst fit to the observed data. This is not surprising once a Gaussian assumption for low counts is inadequate, as mentioned before. Nonetheless, comparison to the normal model is informative to highlight that inadequate modeling assumptions may lead to misleading results. All integer-valued-based approaches support the conclusion that the beta spectral power at lag 10-time units is causal of the spiking rate at time $t$ and, amongst those, the highest degree of certainty is achieved under the proposed bivariate model. 

In addition to the study of lagged causality that was explored in detail in Figure \ref{fig:delta_selection}, our model also supports contemporaneous causality, i.e., causality between $Y_{1t}$ and $Y_{2t}$. This is captured via the $\rho$ parameter, and its posterior distribution is summarised in Table \ref{tab:poisson_gamma_brain}, alongside $\boldsymbol{\beta}^{(1)},\boldsymbol{\beta}^{(2)},\boldsymbol{\alpha}^{(1)}$, $\boldsymbol{\alpha}^{(2)}$ and $\phi$. The posterior distribution of this contemporaneous effect is centered at $-0.079$ as illustrated to the right, in Figure \ref{fig:post_causality}. Therein, the estimated $\rho$ and $\gamma_1, \ldots, \gamma_{15}$ posteriors are plotted. The peak at $0.63$ is due to $\gamma_{10}$, which is the $\beta$-band lag indicated in Figure \ref{fig:delta_selection}. As expected, its posterior distribution is the furthest from zero in comparison with the remaining. Although the 95\% credible interval for $\rho$ contains zero, we have that the probability of a negative $\rho$ is 75\%, $P(\rho <0) = 0.75$. For this reason, we investigate further the joint effect of $Y_{1t}$ and $Y_{1\,t-10}$ in the spiking rate $\mu_{2t}$.

\begin{minipage}{0.4\textwidth}

\begin{tabular}{cclcll}
\hline
\textbf{Parameter} & \multicolumn{2}{c}{\textbf{Mean (SD)}} & \multicolumn{3}{c}{\textbf{95\% CI}}      \\ \hline
$\beta_0^1$        & \multicolumn{2}{c}{0.424 (0.195)}      & \multicolumn{3}{c}{(0.091,  0.820)}       \\
$\beta_0^2$        & \multicolumn{2}{c}{$-$0.120 (0.237)}   & \multicolumn{3}{c}{($-$0.595,  0.343)} \\
$\beta_1^1$        & \multicolumn{2}{c}{0.028 (0.084)}      & \multicolumn{3}{c}{($-$0.136, 0.196)}        \\
$\beta_1^2$        & \multicolumn{2}{c}{0.382  (0.167)}     & \multicolumn{3}{c}{(0.029, 0.684)}        \\
$\alpha_1^1$       & \multicolumn{2}{c}{0.156 (0.368)}   & \multicolumn{3}{c}{($-$0.555, 0.846)}     \\
$\alpha_1^2$       & \multicolumn{2}{c}{0.043 (0.242)}      & \multicolumn{3}{c}{($-$0.299, 0.582)}     \\ 
$\rho$   & \multicolumn{2}{c}{$-$0.079 (0.116)}      & \multicolumn{3}{c}{($-$0.311, 0.144)}     \\ 
$\phi$  & \multicolumn{2}{c}{0.116 (0.020)}      & \multicolumn{3}{c}{(0.125, 0.204)} \\
\hline
\end{tabular}
\captionof{table}{Summaries of the posterior distributions of $\boldsymbol{\beta}^{(1)},\boldsymbol{\beta}^{(2)},\boldsymbol{\alpha}^{(1)}$, $\boldsymbol{\alpha}^{(2)}$, $\rho$ and $\phi$ fitted to the brain recordings data with the Poisson-Gamma model.}\label{tab:poisson_gamma_brain}
\end{minipage}
\hfill\allowbreak%
\begin{minipage}{0.4\textwidth}
\includegraphics[width=.9\textwidth]{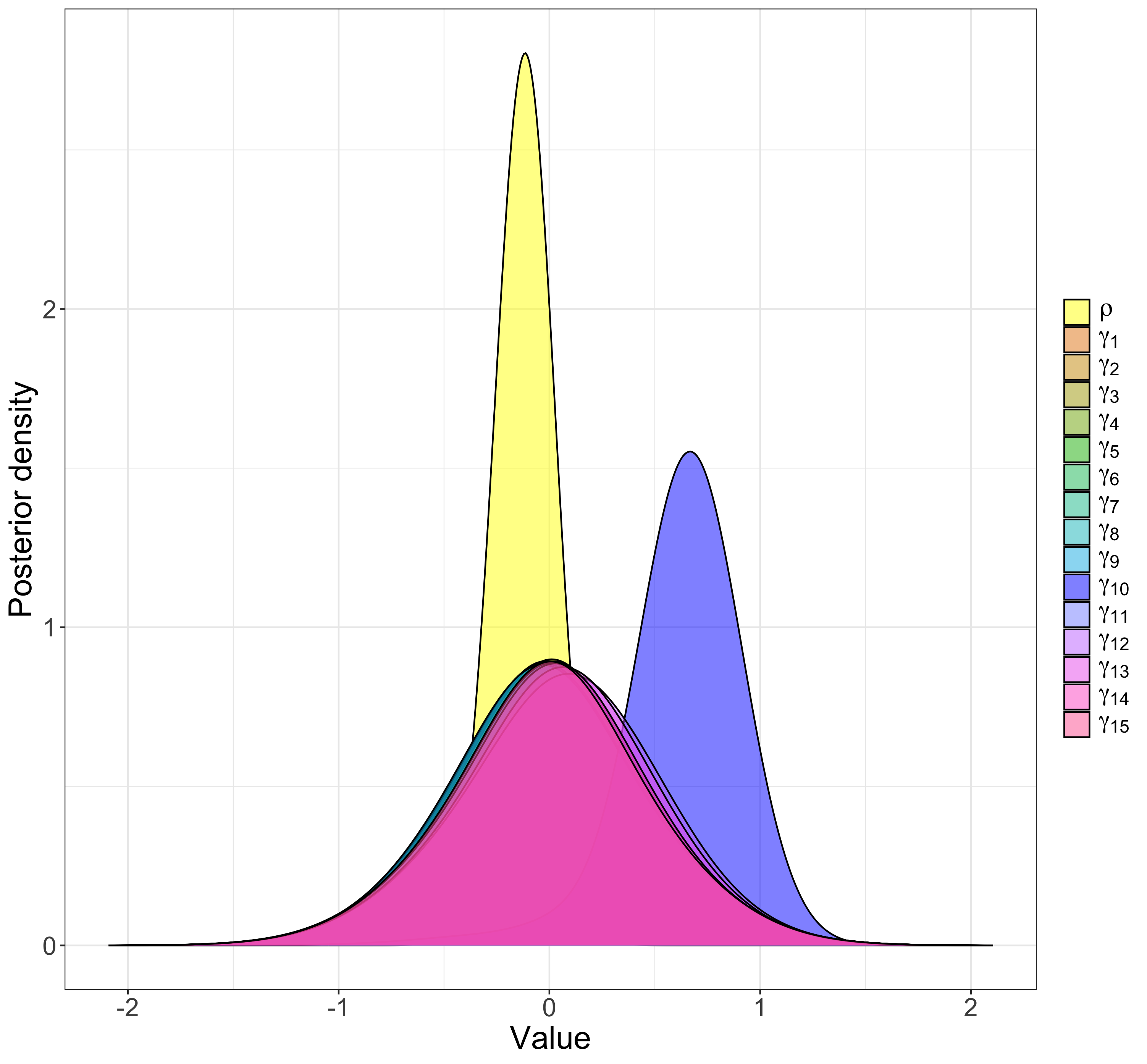}
\captionof{figure}{Posterior distribution of contemporaneous ($\rho$) and lagged ($\gamma_1, \ldots, \gamma_{15}$) causality parameters.}\label{fig:post_causality}
\vspace{1cm}
\end{minipage}

\begin{figure}
    \centering
    \includegraphics[width = 0.85\linewidth]{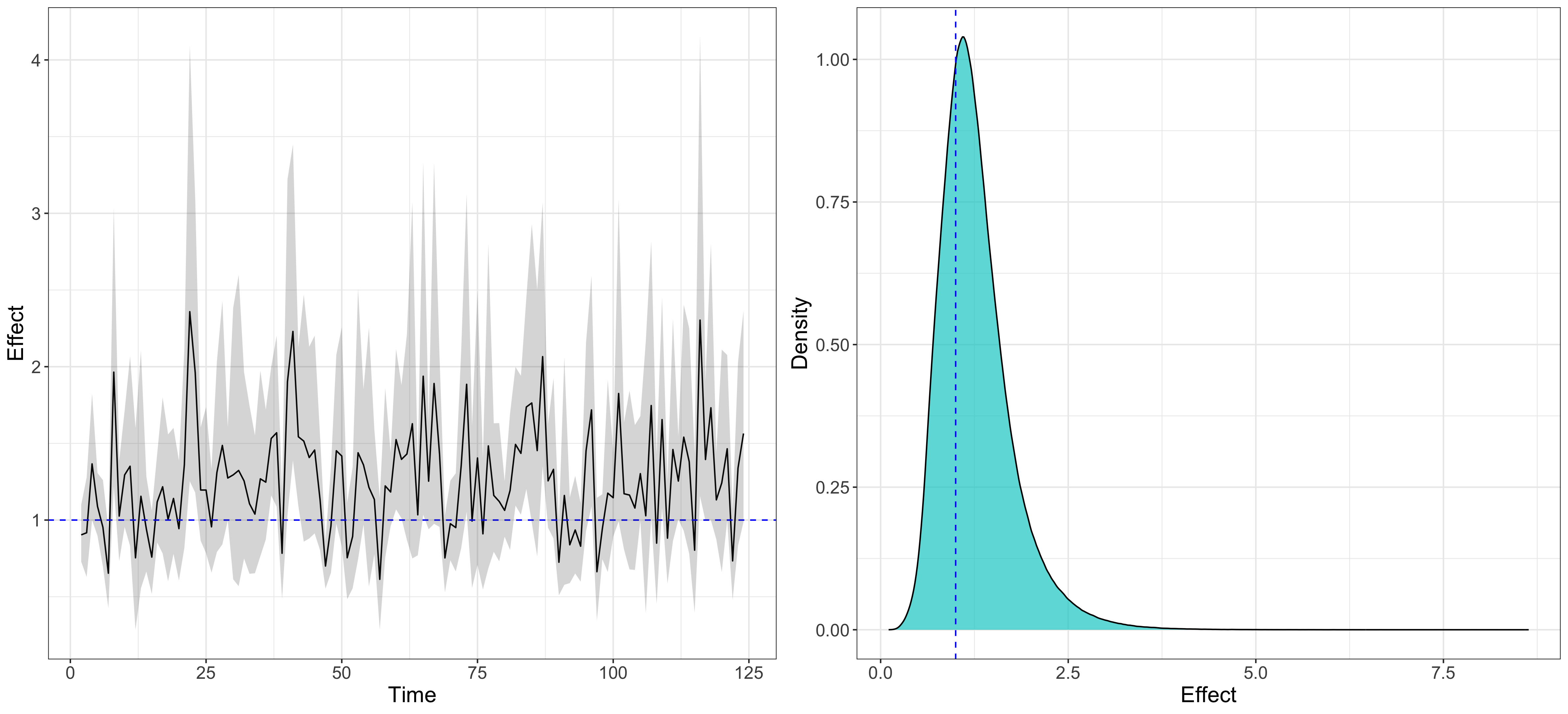}
    \caption{Illustrations of the joint effect of the contemporaneous ($Y_{1t}$) and lagged ($Y_{1,t-1}, Y_{1, t-10}$) LFP effects in the mean spike count at time $t$. To the left, the posterior mean of $\exp(\rho Y_{1t} + \gamma_{10} \log(Y_{1,t-10}))$ is indicated as a solid line at 95\% credible intervals in grey shades. On the right, the same quantity is computed and aggregated across time. }
    \label{fig:fitted_meaneffect}
\end{figure}

Interpretation of the joint effect of the beta spectral power at times $t$ and $t-10$ in the average spiking rate $\mu_{2t}$ should be done preferably in a graphical manner. This is because contemporaneous and lagged effects enter the model in different scales. Their joint effect in $\mu_{2t}$ is measured by $c_t \equiv \exp(\rho Y_{1t} + \gamma_{10} \log(Y_{1\,t-10}))$, where $Y_{1\,t-10}$ is in the log scale. This quantity corresponds to the multiplicative effect from the $\beta$ power (at $t$ and $t-10$) in $\mu^0_{2t} \equiv \exp( \beta_0^{(2)} + \beta_1^{(2)} \log(Y_{2, t-1})+ \alpha_1 \log(\mu^0_{2, t-1}) )$. The notation $\mu^0_{2t}$ is used to indicate the marginal $Y_{2t}$ mean, with no causality. Hence, $\mu_{2t} \equiv \mu^0_{2t} c_t$ and our interest is in the $c_t$ series.  In Figure \ref{fig:fitted_meaneffect}, the left plot displays the posterior mean of $c_t$ for all $t>11$ as a solid line, and a 95\% credible interval is displayed with grey shades. A horizontal dashed line is added for reference as $c_t$ values above one indicate a positive causality between the two series. On the right side of Figure \ref{fig:fitted_meaneffect}, the histogram shows the joint distribution of $c_t$ aggregated over time. This alternative analysis highlights a high posterior probability that the beta spectral power is positively correlated to the average spiking process. In other words, an increase in beta power is associated with an increase in the probability of observing a spike after 300 ms. 

\subsection{Bivariate spike counts data analysis}\label{sec:geo_geo}

\begin{figure}
    \centering
    \includegraphics[width=0.8\linewidth]{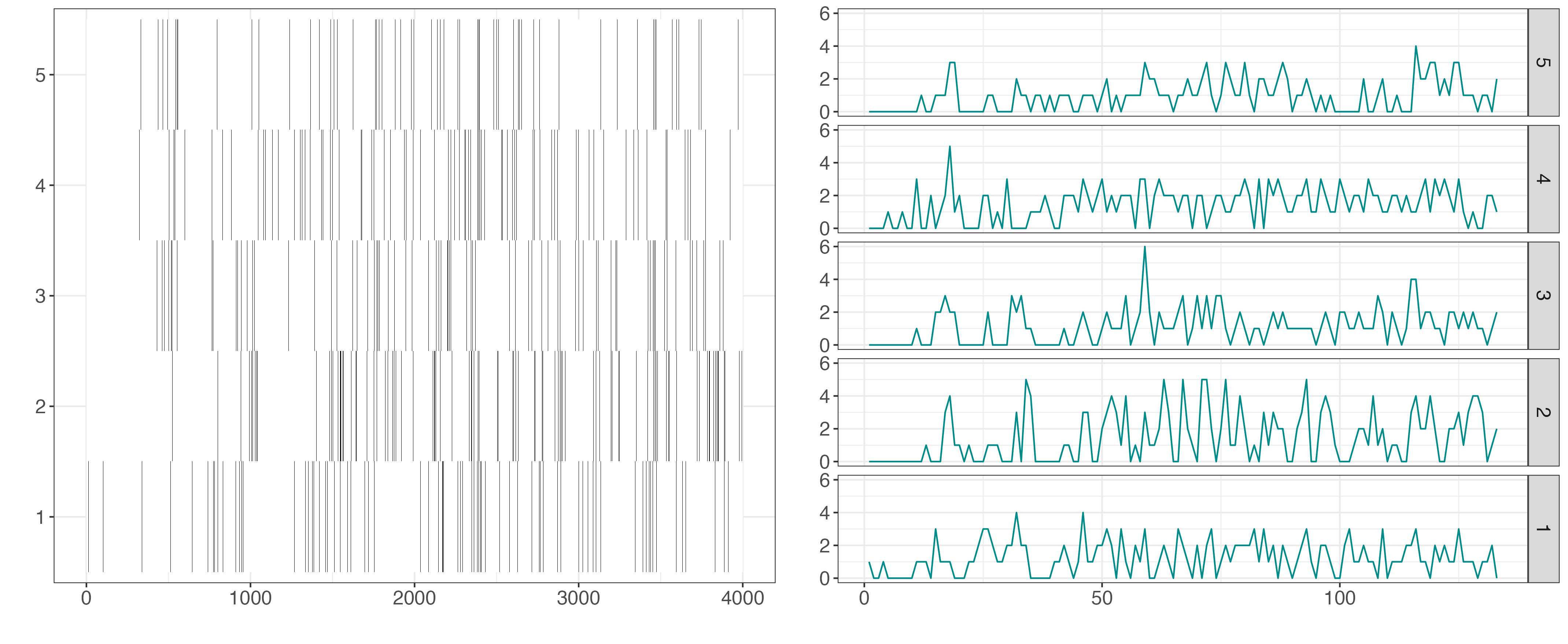}
    \caption{To the left, spike rasters built from activity within each of five electrodes. Count time series displayed on the right is the total number of spikes due to cells in 
    an electrode at non-overlapping windows of 30 points. }
    \label{fig:spikes}
\end{figure}

In this section, we explore the causality of spiking activity between a pair of neurons using the proposed Geometric-Geometric Granger-GLM model. 
As previously noted, count trajectories as computed as the number of spikes within 30 time-point intervals, for each electrode. Our goal is to fit the bivariate model with the structure
\begin{equation}
    \begin{cases}
           Y_{2t}|\mathcal{F}^{(1,2)}_{t-1} \sim \mbox{Geometric}( \mu_{2t} \exp(\rho Y_{1t})),\\
         \mu_{2t} = \exp\left( \beta_0^{(2)} + \sum_{i=1}^p\beta_i^{(2)} \log(Y_{2\,t-i} +1) + \sum_{j=1}^q \alpha_j^{(2)} \log(\mu_{2\,t-1}) + \sum_{l=1}^k \gamma_l \log(Y_{1\,t-l} +1) \right), \\
         Y_{1t}|\mathcal{F}^{(1)}_{t-1} \sim \mbox{Geometric}( \mu_{1t}), \\
         \mu_{1t} =   \exp\left( \beta_0^{(1)} + \sum_{i=1}^r\beta_i^{(1)} \log(Y_{1\,t-i} + 1) + \sum_{j=1}^s \alpha_j^{(1)} \log(\mu_{1\,t-1}) \right), 
     \end{cases}
\end{equation}\label{eq:geogeo}
for all possible ${\boldsymbol{Y}^l_{1}, \boldsymbol{Y}^m_{2}}$ where $l\neq m, l,m \in \{1, \ldots, 5\}$. In (\ref{eq:geogeo}), the geometric distributions are parameterized by the mean. Within each electrode, spiking activity due to five different neurons is measured separately, but we consider neuron aggregations in this analysis due to a very low level of activity from some cells. In Figure \ref{fig:spikes}, electrode-specific rasters are shown with vertical lines indicating a spike at time $t$ is observed for at least one of the neurons measured therein. To the right, count trajectories are constructed for each electrode by summing over non-overlapping windows of size 30. 

In what follows, we examine combinations of spike count trajectories presented on the right side of Figure \ref{fig:spikes} using the Geometric-Geometric model (\ref{eq:geogeo}). Our goal is to evaluate the following hypothesis: \textit{(1): Can we identify synchronous activity between all electrode pairs?} Synchrony is expected as all electrodes are located in the same brain region and thus likely to have some common features in their signal. Secondly, \textit{(2): Although synchrony is anticipated, is there any evidence of lagged causality?} In the presence of a delay or a time lag, one may infer that the spiking activity in one neuron Granger-causes the activity in another neuron, with the effect being either an excitation or inhibition of spiking activity. 

The likelihood ratio test described in section \ref{sec:GC} allows a formal assessment of these hypotheses which is conducted as follows. Suppose that model (\ref{eq:geogeo}) is fitted to a pair ${\boldsymbol{Y}^{\ell}_{1}, \boldsymbol{Y}^m_{2}}$ with arbitrary $p, q, r, s$ and $k=0$. We test for synchronous causality between series $\ell,m$ (in the $l \rightarrow m$ direction) with the set of hypothesis $\mathcal{H}_0: \rho = 0$ versus $\mathcal{H}_1: \rho \neq 0$. As mentioned previously, this involves fitting the model under $\mathcal{H}_1$ and $\mathcal{H}_0$ via maximum likelihood and computing the $LR$ statistic. The asymptotic null distribution of this $LR$, a $\mathcal{X}^2_{1}$, is then used to obtain the $p-$value. The frequentist approach is chosen in this part of our analysis as a way to provide a computationally efficient exploration of all $\ell,m$ combinations. 

\begin{figure}
    \centering
    \includegraphics[width =0.8\linewidth]{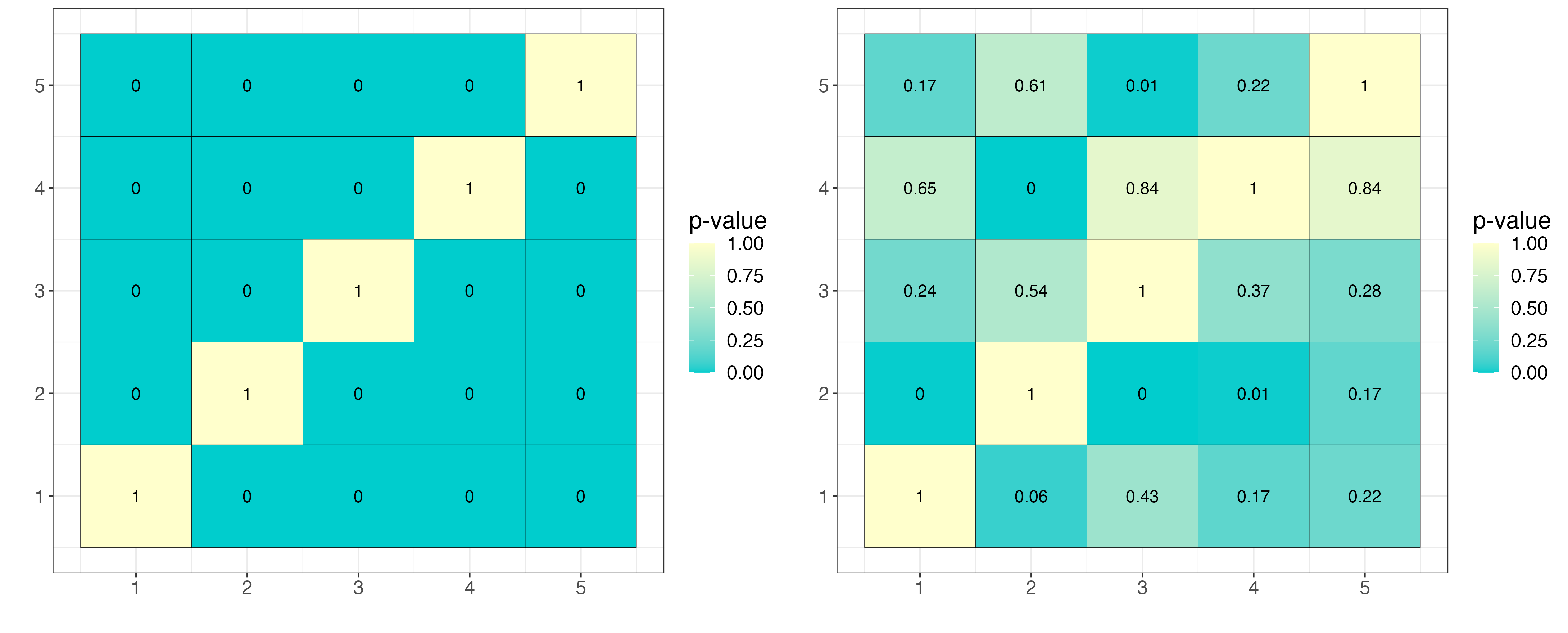}
    \caption{Heatmaps displaying $p-$values obtained from likelihood ratio tests of synchronous causality $\mathcal{H}_0: \rho = 0$ versus $\mathcal{H}_1: \rho \neq 0$ on the left, and lag-one causality $\mathcal{H}_0: \gamma_1 = 0$ versus $\mathcal{H}_1: \gamma_1 \neq 0$ on the right.  On the horizontal axis, the index of the electrode-specific trajectory deemed causal ($Y_1$) of $Y_2$ which is indicated on the vertical axis (i.e., tests are in the direction column causes row).}
    \label{fig:spike_heatmap}
\end{figure}

Results are displayed in Figure \ref{fig:spike_heatmap} in the form of a heatmap. Row and column numbers indicate the pair of neuronal spike count trajectories used to perform the test with colors representing the magnitude of the LR test p-value. The horizontal axis indicates $\boldsymbol{Y}_1$, the electrode fitted as being causal of $\boldsymbol{Y}_2$, which is on the vertical axis. On the left-side heatmap, the hypothesis $\mathcal{H}_0: \rho = 0$ versus $\mathcal{H}_1: \rho \neq 0$ is tested for models fitted with $p=q=r=s=1$. P-values near zero are obtained for all electrode combinations, confirming that synchronous activity holds for all pairs, in any causality order. The right-side heatmap shows the results of repeating the analysis for the hypothesis $\mathcal{H}_0: \gamma_1 = 0$ versus $\mathcal{H}_1: \gamma_1 \neq 0$. In this case, $\mathcal{H}_0$ and $\mathcal{H}_1$ models include contemporaneous and lag-one causality and take $p=q=r=s$ as before. In other words, this $LR$ test stipulation compares between the nested models that include $Y_{1t}$ or both $Y_{1t}$ and $Y_{1,t-1}$ to explain $\mu_{2t}$. 

Important comments to make regarding this second test are the following. First, results now are not uniform for all electrode pairs, indicating that a lagged relationship is relevant in some, but not all cases. Secondly, the matrix is no longer symmetric because results change depending on the causality direction. For example, take the pair $l=2,m=3$ where $l$ is the row index and $m$ is the column index. This test addresses the question of whether or not the prior spiking
activity of electrode 3 is causal to the future spiking activity of electrode 2. This returns a p-value near zero, so we can reject $\mathcal{H}_0$ and conclude that there is evidence of lagged causality. On the other hand, when $l=3,m=2$, analysis of the same pair in the ${Y}^2 \rightarrow Y^3$ direction produces $p = 0.54$. Another electrode pair for which the test's conclusion changes depending on direction is $\{3, 5\}$. 

\begin{figure}
    \centering
    \includegraphics[width = 0.8\linewidth]{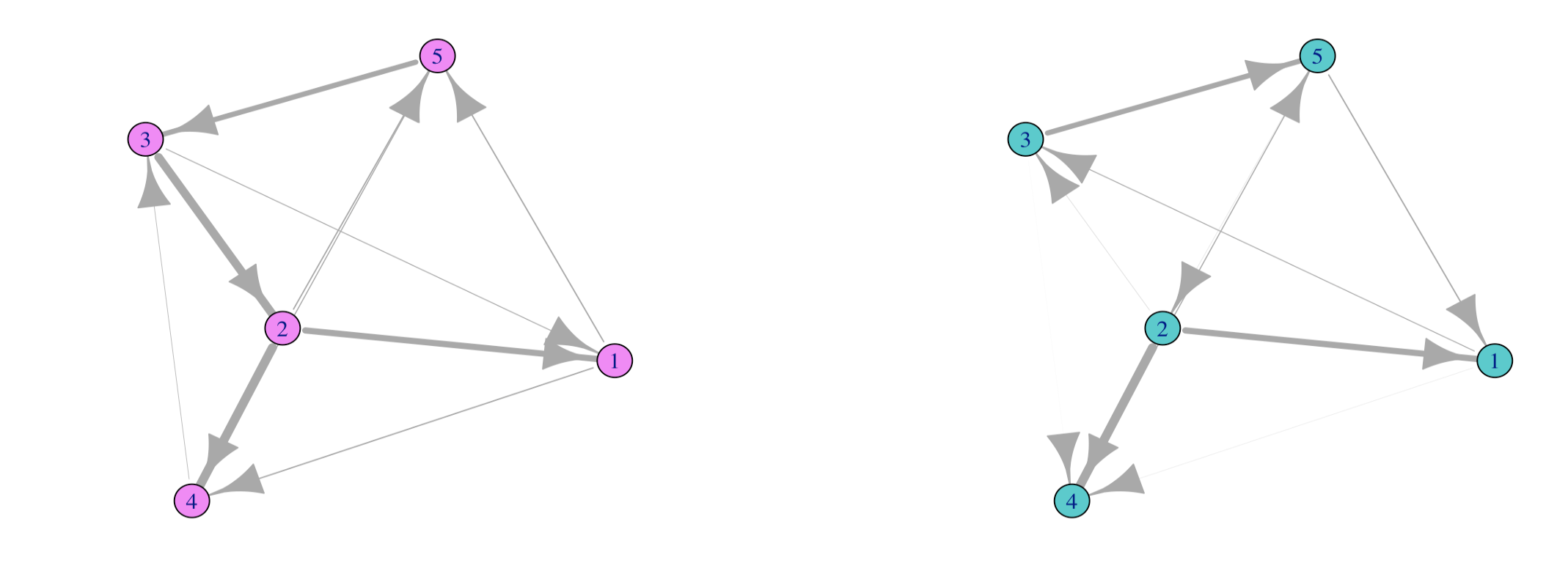}
    \caption{Network visualization of the likelihood ratio tests of  $\mathcal{H}_0: \gamma_1 = 0$ versus $\mathcal{H}_1: \gamma_1 \neq 0$. This is an alternative visualization of the right-hand-side heatmap of Figure \ref{fig:spike_heatmap}, where edge widths are proportional to the LR test statistic. The upper triangle of this matrix is represented by the left-side network, and the lower triangle is on the right.}
    \label{fig:network}
\end{figure}

We conclude this section by highlighting that this type of analysis is scalable to larger numbers of cells since the model fit under the frequentist approach requires a small computational effort. Another useful representation of the same information is with a network structure. In Figure \ref{fig:network}, this is illustrated for the $\mathcal{H_0}: \gamma_1 = 0$ versus $\mathcal{H}_1: \gamma_1 \neq 0$ test results. Networks are constructed with edge widths proportional to the test statistics from the upper triangle of the matrix (\ref{fig:spike_heatmap}) on the left, and lower triangle on the right. 

\section{Discussion of the Results}\label{sec:discussion}

We introduced novel Granger causality tests to study multimodal brain activity data. We explored causality between the spectral power of LFPs and the number of spikes, and also pairwise dependencies between spike trajectories from five different electrodes. These studies motivated us to develop novel Granger tests to handle mixed types of data. To do that, a family of bivariate generalized linear models time series was introduced. Inferences based on frequentist and Bayesian approaches were proposed, with a focus on the latter. A spike and slab prior was considered to enable us to perform Granger order selection, an important problem in such an area. Critical insights into the causal relationship between the rat spiking
activity and LFP spectral power were obtained. Performing the same task based on other methods revealed that causality order selection can be misleading and show increased uncertainty under alternative approaches. Hence, having a methodology that is tailored to the bivariate order selection study is paramount. We believe that this novel analytical tool can be very useful for this area of emerging interest in neuroscience. At the same time, our method can be applied in other areas. We hope to report such scalability in a future paper. We have initial findings that show the method works quite well in other areas like epidemiology. Other points deserving future investigation are: (i) to establish theoretical properties of the bivariate Granger-GLM and asymptotics of the frequentist approach based on maximum likelihood estimation; (ii) higher-order model formulation allowing for more than two time series; (iii) incorporation of covariates to handle non-stationarity; (iv) joint order selection/estimation of $(p,q,r,s,k)$.

\end{document}